\documentclass[modern,preprintnumbers]{aastex62}
\usepackage{fullpage}
\usepackage{graphicx}
\usepackage{amsmath}
\usepackage{amssymb}
\usepackage{color}
\usepackage[caption=false]{subfig}

\received{May 9, 2019}
\accepted{July 9, 2020}
\published{September 1, 2020}
\submitjournal{Astrophys. J.}

\begin{document}

\title{Testing $\Lambda$CDM With Dwarf Galaxy Morphology}

\author{Weishuang Linda Xu}
\affil{Harvard University, 17 Oxford St, Cambridge, MA, 02139}
\email{weishuangxu@g.harvard.edu}

\author{Lisa Randall}
\affil{Harvard University, 17 Oxford St, Cambridge, MA, 02139}
\email{randall@physics.harvard.edu}

\begin{abstract}
The leading tensions to the collisionless cold dark matter (CDM) paradigm are the ``small-scale controversies", discrepancies between observations at the dwarf-galactic scale and their simulational counterparts.  In this work we consider methods to infer 3D morphological information on Local Group dwarf spheroidals, and test the fitness of CDM+hydrodynamics simulations to the observed galaxy shapes.  We find that the subpopulation of dwarf galaxies with mass-to-light ratio $\gtrsim 100 M_\odot/L_\odot$ reflects an oblate morphology.  This is discrepant with the dwarf galaxies with mass-to-light ratio $\lesssim 100 M_\odot/L_\odot$, which reflect prolate morphologies, as well as simulations of CDM-sourced {bright isolated} galaxies which are explicitly prolate. Although more simulations and data are called for, if evidence of oblate pressure-supported stellar distributions persists {in observed galaxies while being absent from simulations}, we argue that an underlying oblate non-CDM dark matter halo may be required,  and present this as motivation for future studies.
\end{abstract}

\section{Introduction}

The minimal cold dark matter (CDM) paradigm, which takes  dark matter (DM) to be cold and collisionless, has been quite successful, particularly at high redshift and large scales (\cite{Ade:2015xua}).  However, its particle nature is still very much unknown, and a wide variety of direct and indirect detection experiments are actively searching for evidence of any nongravitational coupling to the visible sector (\cite{TheFermi-LAT:2017vmf,Xiao:2017vys, Angle:2007uj, Amole:2017dex, Aalseth:2012if, Essig:2012yx}). As it stands, the parameter space remains open for many models of interacting dark matter (\cite{ArkaniHamed:2008qn, Sigurdson:2004zp, Preskill:1982cy,Jungman:1995df,Buckley:2009in, Feng:2009mn,Kaplan:2009ag}). Correspondingly, many new ways of testing for the properties of dark matter have yet to be explored. 

At small astrophysical scales in particular, dwarf galaxies have significant promise as an arena for investigating DM interactions. They are strongly DM-dominated systems offering both a uniquely clean testing ground of the halo gravitational potential and a probe of the smallest scales of the matter power spectrum. Despite having low luminosities, many candidates have been observed due to their relative proximity.  In fact, the current leading potential challenges to CDM come from discrepancies between simulations and observations at the galactic and dwarf-galactic scale.   Specifically, the ``core-cusp'' (\cite{Bullock:2017xww,Spergel:1999mh,Walker:2011zu}), ``missing satellite'' (\cite{Klypin:1999uc,Weinberg:2013aya}), and ``too big to fail'' (\cite{BoylanKolchin:2011de, Papastergis:2014aba}) problems  have prompted speculation that observed dwarf galaxies are less numerous, less massive, and less centrally dense than their CDM simulation counterparts.  

Ultimately the goal is to see how well the CDM paradigm fits with the observed data as we understand it.  While it is unclear that new physics is necessary to resolve such discrepancies (\cite{Brooks:2012vi, Read:2017lvq,Arraki:2012bu,2018MNRAS.474.1398G}), these tensions have motivated a lively collection of self-interacting dark matter models (\cite{Buckley:2009in,Fan:2013yva, Cyr-Racine:2015ihg,Bode:2000gq,Boehm:2001hm,Moore:1999nt}), that propose as a solution non-minimal dark sectors which couple to the visible sector only via gravity. These models would therefore be inaccessible to direct and indirect detection but would induce modifications to structure formation at small scales. 

Detailed efforts have been invested in characterizing the stellar populations and kinematic parameters of the classical dwarf galaxies, in the service of understanding the surrounding halo structure (\cite{2009MNRAS.394L.102L,  2009ApJ...704.1274W, 2013RAA....13..517E,2012ApJ...756L...2A,Martin:2008wj,2011ApJ...733L..46W, 2015MNRAS.453.1047P,2018MNRAS.476.2168M,Walker:2008ax,Hayashi:2016kcy}).  But although there has been a good deal of attention directed towards probing different mass profiles, relatively little exploration has been dedicated to studying dwarf galaxy shapes.  In part this omission is due to the reality that we observe only 2-D projections, and full morphological information is elusive without precise 3-D spatial and kinematic data for individual stars. 

In this paper, we consider the morphologies of dwarf spheroidals (dsphs) in and around the Local Group (LG). In particular, we investigate their ellipticities and associated 3-D shapes, and whether they might be prolate or oblate.  We then test the ability of CDM-sourced dwarf galaxies, as represented by hydro-intensive simulations, to fit the observed morphologies.  Dsphs were chosen in particular since they are the most common and are typically found to have very little gas, such that when in equilibrium their stellar distributions are determined entirely by the underlying gravitational potential. Simulations of CDM have unilaterally produced halos that are prolate (where the two shorter axes are closer to each other than to the longer axis) (\cite{1996ApJ...462..563N, 2012AJ....144....4M}), and as dsphs have been found to be relatively isotropic in stellar velocity dispersion and not rotationally supported (\cite{ Walker:2008ax, 2013RAA....13..517E, 2015ApJ...808..158B,annurev.astro.36.1.435}), we expect the stellar distributions of dsphs sourced from CDM haloes should be likewise prolate.  In this paper we consider methods to infer the 3-D baryonic distributions of observed dwarf spheroidals in the local group, catalogued in Refs. \cite{2012AJ....144....4M,2018arXiv180606889M}, to test this hypothesis.  In particular, we compare these results to the shapes of galaxies from a suite of 14 CDM-sourced dwarf galaxies with star formation from the Feedback In Realistic Environments (FIRE) project (\cite{Fitts:2016usl, Hopkins:2017ycn}).    

We find that there is currently a discrepancy between the morphologies of LG dwarf galaxies with mass-to-light ratio $> 100 M_\odot/L_\odot$ and those of the FIRE dwarf galaxies, and this subpopulation of LG dwarf galaxies is consistent with a more oblate morphology not reflected in simulations. We present this as a template and motivation for more statistically robust future studies.  
A mismatch between the CDM picture and the physical observations can point to a deficiency in our understanding of either the observed or simulational data, or it might challenge the CDM paradigm and motivate models that produce galaxies with a different morphology.

In section~\ref{GR} we describe our geometric assumptions on dsphs stellar profiles, and our methods for inferring their 3-D distributions. We propose using correlations between ellipticity and surface brightness, as well as ellipticity and line-of-sight velocity dispersion, as a measure for the oblateness or prolateness of observed dwarf galaxies. Similar relations have been considered in the context of larger galaxies in Refs. \cite{Vincent:2005xr, Bosch:2008ts}, though the comparative richness of internal structure in that regime makes this relationship more subtle.  In section~\ref{data} we present the observational and simulational data used in our analysis. In sections~\ref{up} and~\ref{down} we summarize results from projecting simulations and deprojecting observations respectively. We find that the population of observed LG dwarf galaxies with mass-to-light ratio $>100 M_\odot/L_\odot$ (``dim" dwarfs) are correlated in observed ellipticity and central surface brightness (stellar velocity dispersion), consistent with oblateness, and those with mass to light ratio $< 100 M_\odot/L_\odot$ (``bright" dwarfs) exhibit anti-correlation consistent with prolate morphologies. Projecting simulated {isolated} galaxies over a distribution of observing angles gives anti-correlation consistent with the bright dsphs and significantly discrepant with the dim subpopulation, {indicating that the latter may be a morphologically distinct subgroup}. In section~\ref{tests} we test the robustness of our results by subsampling both the observed and simulated data, varying the choice of mass-to-light threshold, and relaxing assumptions on observation angle distributions. We show that this discrepancy holds under these tests. In section~\ref{bary} we investigate implications for DM and baryon distribution axis ratios, and consider the specific case of dissipative dark matter. Section~\ref{conc} explores avenues of future work and presents our conclusions.

\section{Geometric Remarks}\label{GR}

For the present, observational data of dwarf galaxies is constrained to on-sky-projected positional and line-of-sight velocity information for its individual stars.  Unfortunately the remaining line-of-sight spatial and on-sky proper motion data remain out of reach at the typical distances of dwarf satellites, which is of order 30 kpc in the closest cases.  The recent Gaia DR2 (\cite{2016A&A...595A...1G})  provides extremely precise astrometric measurements of parallax and proper motion up to uncertainties of 0.04 mas and 0.06 mas/yr respectively for the brightest stars. At dwarf-galactic distances however,  this translates to line of sight distance resolved up to  $\sim 20$ kpc and on-sky velocities up to $\sim 15$ km/s in the maximally optimistic cases.  As these dwarfs are typically of half-light radius $\sim 0.1 $ kpc with line-of-sight stellar velocity dispersion $\sim 5$ km/s, the current best measurements are far from able to resolve the full spatial and kinematics directly. Thus we will focus on what one can hope to infer from the correlation between currently accessible observables and geometry.

 As is done in Refs. \cite{Walker:2011zu,2009ApJ...704.1274W,2012ApJ...756L...2A} , we will model the stellar distribution with a Plummer profile (\cite{1911MNRAS..71..460P}), and assume that mass follows light.  Similar results may be obtained for King or Exponential profiles (\cite{1962AJ.....67..471K}). We include a triaxiality ($a\ge b \ge c$)  with ellipsoidal isophotes 

\begin{equation}
\rho_*  = \rho_0 (1+ \frac{x^2}{a^2} + \frac{y^2}{b^2} + \frac{z^2}{c^2} )^{-5/2}
\end{equation}

We fix the observer to be at some angle    \begin{equation}\hat z'  \equiv  \sin\theta \cos \phi \hat x + \sin \theta \sin \phi \hat y + \cos \theta \hat z  \end{equation}  

The observer then defines a right-handed coordinate system $(x',y',z')$ where $x'$ and $y'$ parametrize the plane of projection with a remaining rotational degree of freedom

 \begin{equation} \begin{bmatrix} x' \\ y' \\z' \end{bmatrix} =   \begin{bmatrix} \cos \omega & -\sin\omega &0 \\ \sin\omega & \cos\omega & 0 \\ 0 & 0& 1\end{bmatrix}   \begin{bmatrix} \sin\phi  & -\cos\phi &0 \\ \cos\theta\cos\phi & \cos\theta\sin\phi & -\sin\theta \\ \sin\theta \cos\phi & \sin\theta\sin\phi & \cos\theta \end{bmatrix} \begin{bmatrix}  x\\ y\\z \end{bmatrix}   \end{equation}

To obtain the surface brightness profile $S_*(x,y)$, we can then rewrite the density profile in the new coordinates and integrate along the line of sight. We choose $\omega$ such that $x'$ and $y'$ align with the semi-major and semi-minor axis of the projected ellipse.  the surface brightness is then given by

\begin{align} S_* (x,' y') & = \int dz' \rho_* = \frac{4}{3} \rho_0 \frac{abc}{\sqrt{\alpha \beta - \gamma^2}} \nonumber\\
 & \times \left( 1 + \frac{x'^2}{ \left( \frac{1}{2}(\alpha + \beta) - \frac{1}{2} \sqrt{4\gamma^2 + (\alpha-\beta)^2}\right)} 
+   \frac{y'^2}{ \left(\frac{1}{2}(\alpha + \beta)  + \frac{1}{2} \sqrt{4\gamma^2 + (\alpha-\beta)^2} \right)}  \right)^{-2}
\end{align}
 
where $\alpha, \beta, \gamma$ are parameters dependent on the principal axes and observer angles, expressed as \begin{equation}\begin{cases}  \alpha = a^2 \cos^2\theta \cos^2\phi + b^2 \cos^2\theta \sin^2\phi  + c^2 \sin^2\theta \\
\beta =  a^2 \sin^2\phi + b^2 \cos^2\phi \\
\gamma = (a^2 - b^2) \cos \theta \sin \phi \cos \phi 
 \end{cases}
 \end{equation}

Thus we can write down  projected observables (ellipticity $\epsilon$ and central surface brightness $\Sigma_* = S_*(0,0)$)  explicity in terms of observer angle and intrinsic parameters (principal axis ratios $\frac{b}{a}, \frac{c}{a}$ and central density $\rho_0$)
\begin{equation}\begin{cases}  \epsilon & = 1- \sqrt{1- \frac{2\sqrt{4\gamma^2 +(\alpha-\beta)^2}}{\alpha + \beta + \sqrt{4\gamma^2 +(\alpha-\beta)^2}}} \\
\Sigma_* & =  \frac{4}{3} \rho_0 \frac{abc}{\sqrt{\alpha \beta - \gamma^2}}
 \end{cases} \end{equation}

\begin{figure}[h!!]
    \centering
    \subfloat[oblate  ($b /a =1, c/a=0.4$)]{
        \includegraphics[width=0.45\textwidth]{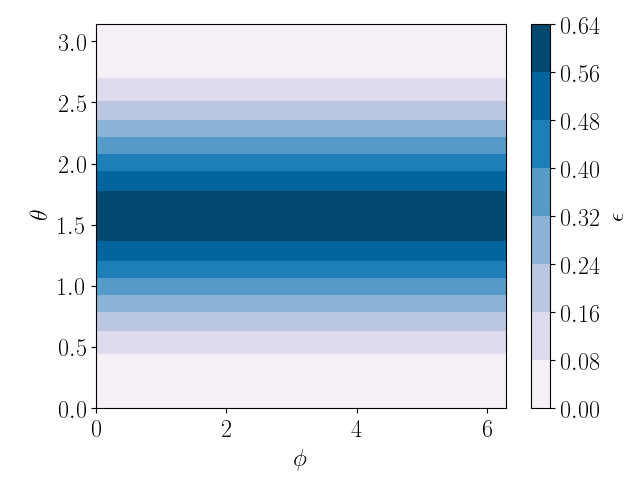}
        \label{fig:obs}}
    ~ 
    \subfloat[prolate ($ b/a= c/a=0.4$)]{
        \includegraphics[width=0.45\textwidth]{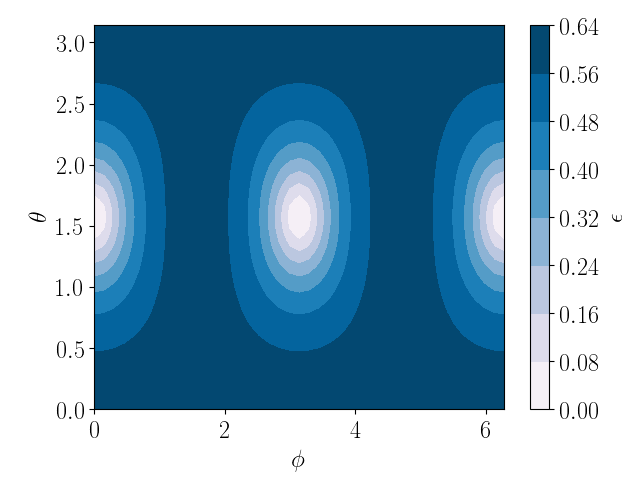}
        \label{fig:pro}}
    
    \subfloat[triaxial  ($b/a =0.7, c/a=0.4$)]{
        \includegraphics[width=0.45\textwidth]{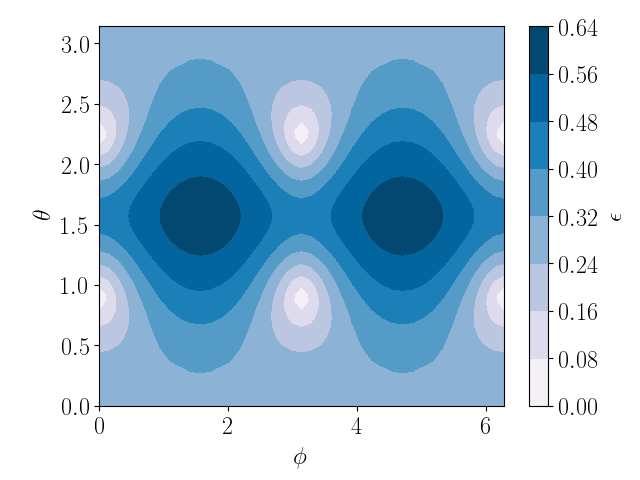}
        \label{fig:tri}}
    \caption{Contour map of projected ellipticity in  $(\phi , \theta)$ space for various ellipsoids. Note that equal area in $(\phi,\theta)$ space does not imply equal area on the sphere-- in particular the top and bottom lines of these plots are contracted to a point }\label{proj_ellips_theory}
\end{figure}

Figure~\ref{proj_ellips_theory} shows contour maps of projected ellipticity in $(\phi , \theta)$ space for an oblate ($b /a =1, c/a=0.4$), prolate ($ b/a= c/a=0.4$), and  triaxial  ($b/a =0.7, c/a=0.4$), object respectively, for reference.

In fact, in the special cases of oblate $(a=b>c)$ and prolate$(a>b=c)$ bodies, these relations reduce to the more simplified forms 

 \begin{equation}\begin{cases} \epsilon^{obl}& = 1-  \sqrt{\cos^2\theta + \frac{c^2}{a^2} \sin^2 \theta }\\
\Sigma^{obl}_{*} & = \frac{4}{3}  \rho_0 c / \left( 1 - \epsilon^{obl} \right) \\ 
\\
 \epsilon^{pro} & = 1-  \displaystyle{\frac{1}{\sqrt{\frac{a^2}{b^2}(\sin^2\phi + \cos^2\theta \cos^2\phi) + (\sin^2\theta \cos^2\phi)}}}\\
\Sigma^{pro}_{*} & = \frac{4}{3} \rho_0 a \left( 1 - \epsilon^{pro} \right) \\
 \end{cases}  \end{equation}

\begin{figure}[h!!]
    \centering
        \includegraphics[width=\textwidth]{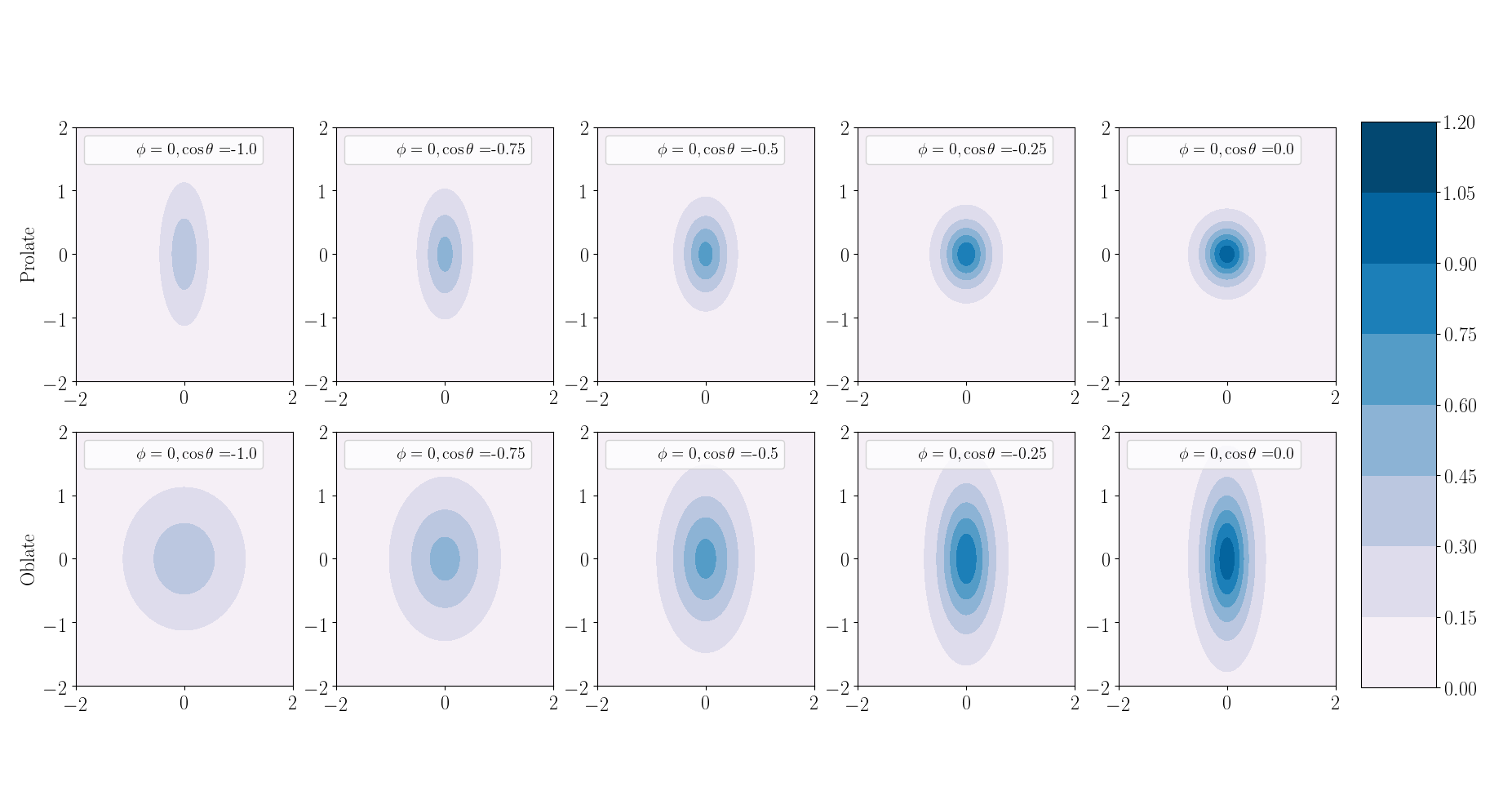}
    \caption{Surface density contours for different projections of a prolate galaxy with $b/a= c/a = 0.4$ (top) and an oblate galaxy $b/a =1,  c/a =0.4$ (bottom). As shown, a lower observed ellipticity is correlated with a higher central surface brightness for prolate galaxies, and the opposite is true for oblate galaxies.} 
      \label{corrpic}
\end{figure}

In particular, we observe that the central surface brightness of a prolate galaxy is anti-correlated with its projected ellipticity, while the opposite is true of oblate galaxies. This is understood intuitively as observing a lower-ellipticity projection of a prolate galaxy implies that the observer is ``looking through" the long axis, while observing the same for an oblate galaxy implies that the short axis lies on the line-of-sight.  This is illustrated in Fig.~\ref{corrpic}, and looking for correlations and anti-correlations between central surface brightness and ellipticity in LG dwarf satellites will allow us to distinguish between oblate and prolate morphologies.

In addition to the central surface brightness we will also consider the line-of-sight stellar velocity dispersion $\sigma_*$ and its correlation with projected ellipticity.  For this work we will assume that the velocity dispersion tensor $\sigma^2_{ij}$ is diagonal along the principal axes of the galaxy and has negligible spatial dependence within the half-light radius, and we will also assume that the velocity dispersion along a larger principal axis is systematically larger. The latter assumption is also reflected in the dwarf galaxy simulations we consider (see Section \ref{data}),  whose full kinematic and spatial data is accessible.  Previous studies bounding stellar velocity dispersion anisotropies in classical dwarf galaxies (\cite{Walker:2008ax,Walker:2007ju,2009MNRAS.394L.102L,Lokas:2001gy}) have assumed that the velocity dispersions are the same in all radial directions.  As we are explicitly looking for violations of this assumption, these bounds do not apply here.   

Thus, given the observing angle $(\phi, \theta)$, the line-of-sight stellar velocity dispersion is given simply by  \begin{equation}\sigma_*^2 = \sin^2\theta \cos^2 \phi \; \sigma^2_{a} + \sin^2 \theta \sin^2\phi \; \sigma^2_b + \cos^2 \theta  \; \sigma^2_c \end{equation}

Where $\sigma_{a,b,c}$ are the stellar velocity dispersions along each of the principal axes. In this way, we expect a similar type of (anti-) correlation between observed ellipticity and measured stellar velocity dispersion for oblate (prolate) dwarf galaxies:  a prolate dwarf projected to have small ellipticity has its semi-major axis along the line-of-sight and will measure a correspondingly large stellar velocity dispersion; the opposite is true for an oblate dwarf.

Naturally, the true relationship between morphology and these correlations is less clean-cut. Firstly, these galaxies do not have perfect axial symmetry -- they are not absolutely prolate or oblate -- and these correlations will degrade with significant triaxiality. Secondly, instead of many projections of a single galaxy, we are observing one projection each of many different galaxies -- each with an intrinsically different shape and luminosity. And thirdly, we have access only to finitely many galaxies at this scale.  

The third point is unfortunately a limitation of nature. To address the first and second point, however, we use simulated galaxy data as a measure for the expected amount of triaxiality and intrinsic dispersion between galaxies in size and brightness. The guiding principle is this: current CDM simulations both with and without baryons, indicate that dwarf spheroidals are prolate, so it is informative to test this prediction against observable data. Any deviation from this expectation -- whether by a subpopulation of oblate or general mix of prolate and oblate shapes -- {warrants a closer monitoring, to check if any discrepancy persists as the simulation and observation data evolve.}

Using the relation between intrinsic ($\frac{b}{a}, \frac{c}{a}$) and projected ($\epsilon$) geometry, we can begin to make morphology comparisons between simulated galaxies and observed ones,  if a distribution of observing angles is assumed. This can be done in one of two ways: projecting down the simulated galaxies to the plane of observation or deprojecting up the observed galaxies into a distribution of probable axis ratios $(b/a, c/a)$.

In the absence of more informative priors, we will assume a distribution of observing angles uniformly sampled in $(\phi, \cos \theta )$ space: 
 \begin{equation} \phi = 2\pi u \qquad \theta = \cos^{-1}(2v-1)  \end{equation}
for variables $u, v$ uniformly distributed on $[0,1]$. In practice, since many of these dwarfs are satellites of Andromeda or the Milky Way, they experience tidal locking and the angle of observation is in fact not uniformly random. We investigate the effects of relaxing this assumption in section~\ref{tests}.

\section{Overview of Observational and Simulational Data}\label{data}

In this section we give a brief description of both the observational and simulational data used in this paper. Further details can be found in Refs. \cite{2012AJ....144....4M, Fitts:2016usl, Hopkins:2017ycn,2018arXiv180606891M,2018arXiv180606889M}.

We are using the observational data compiled by McConnachie (\cite{2012AJ....144....4M}) as well as updated parameters from the MegaCam survey in Ref. \cite{2018arXiv180606889M} as a measure of dwarf galaxy properties.  Of these catalogs, we take only the dwarfs that have been classified as spheroidal galaxies and for which a stellar velocity dispersion has been resolved. We also remove galaxies with strong evidence of irregular structure and tidal stripping (\cite{2018arXiv180606889M, Koch:2008wx, Fellhauer:2006jr,Okamoto:2008td}) as we are considering dwarfs which can be well-modeled with ellipsoidals. This amounts to 18 satellites of the Milky Way, 6 satellites of Andromeda, and 2 isolated dwarfs in the Local Group.  Ref.~\cite{2018arXiv180606891M} reports structural parameters for fits to exponential, Plummer, King, and Sersic models, though only Milky Way satellites are considered. In contrast, Ref.~\cite{2012AJ....144....4M} provides structural parameters for Milky Way, Andromeda, and field dwarfs for an inconsistent set of profiles.  We will use the Plummer structural parameters when possible in this work, though the primary parameter of interest is the ellipticity $\epsilon$, and reported parameters demonstrate that it are not particularly sensitive to choice of profile. These dwarfs span from 30 - 900 kpc in heliocentric distance,  from $10^3  - 10^7 M_\odot$  in stellar masses, from $10^6 - 10^8 M_\odot$ in dynamical masses, and reflect a variety of formation histories. While the two catalogs are in excellent agreement with each other, for galaxies that appear in both sets (the Milky Way satellites) we take parameters from the more recent analysis. Table~\ref{tab:obsgals} lists the galaxies, parameters, and associated references used in this study.

\begin{table}[]
    \centering
    \begin{tabular}{l c c c c c r}
    \hline
    \hline
        Name &  Group &    $\epsilon$ & $\sigma_*$  & $\Sigma_{*}$  &  $M/L_V$ &   Ref. \\
        &   &    &  $[\text{km s}^{-1}]$ & $[\text{mag arcsec}^{-1}]$ &   $[M_\odot/L_{\odot}]$ &  \\
        \hline
        \hline
        Segue I & MW &  $0.33 \pm 0.10$ & $3.9 \pm 0.8$ & $28.06^{+ 1.01}_{- 0.98}$ & 1530  & M18b, S11 \\
        Segue II & MW & $0.22 \pm 0.07$ & $3.4\pm 2.5$ & $28.48 \pm 1.06$ & $500$ & M18b, K13a \\
        Willman 1 & MW & $0.47 \pm 0.06$ & $4.3^{+2.3}_{-1.3}$  & $25.87 \pm 0.94$ & 520 & M18b, K13b, W11\\
        ComBer   &  MW & $0.37 \pm 0.05$ & $4.6 \pm 0.8$ & $26.98 \pm 0.37$ & 500 & M18b, K13b, S07\\
        Bootes I  & MW &  $0.30 \pm 0.03$ & $2.4^{+0.9}_{-0.5}$ & $28.38 \pm 0.30$ & 60 & M18b, K11\\
        Draco  & MW &   $0.29 \pm 0.01$ & $9.1\pm 1.2$ & $25.12 \pm 0.07$ & 80  & M18b, K13b \\
        Ursa Minor & MW &  $0.55 \pm 0.01$ & $9.5 \pm 1.2$ & $25.77 \pm 0.08$ & 70 & M18b, K13b\\
        Sculptor  & MW &  $0.33 \pm 0.01$ & $9.2 \pm 1.4$ & $23.29 \pm 0.15$ & 12 & M18b, K13b, W09a \\
        Sextans I  & MW &  $0.30 \pm 0.01$ & $7.9 \pm 1.3$ & $27.22 \pm 0.08$ & 98 & M18b, K13b, W09a\\
        Carina & MW &  $0.36 \pm 0.01$ & $6.6 \pm 1.2$ & $25.35 \pm 0.07$ & 34  & M18b, W09a \\
        Hercules & MW &  $0.69 \pm 0.03$ & $3.7 \pm 0.9$ & $26.82 \pm 0.39$ & 90  & M18b, K13b\\
        Fornax  & MW &  $0.29 \pm 0.01$ & $11.7 \pm 0.9$ & $23.59 \pm 0.16$ & 5.7 & M18b, K13b, W09a\\
        Leo IV  & MW &  $0.17 \pm 0.09$ & $3.3 \pm 1.7$  & $27.80^{+0.53}_{0.50}$ & 145 & M18b, K13b, S07 \\
        CV II  & MW &  $0.40 \pm 0.13$ & $4.6 \pm 1.0$ & $26.50^{+0.68}_{0.63}$ & 230 & M18b, K13b, S07\\
        Leo V & MW &   $0.43 \pm 0.22$ & $3.7^{+2.3}_{-1.4}$  & $23.29^{+0.90}_{0.79}$ & 215 & M18b, W18b\\
        CV I  & MW &  $0.44 \pm 0.03$ & $7.6 \pm 0.4$  & $26.78 \pm 0.13$ & 160 &  M18b, K13b, S07\\
        Leo II & MW &   $0.07 \pm 0.01$ & $6.6 \pm 0.7$  & $24.24 \pm 0.07$ & 13  & M18b, K13b, K07\\
        Leo I & MW &   $0.30 \pm 0.01$ & $9.2 \pm 1.4$  & $22.61 \pm 0.30$ & 4.4  & M18b, K13b\\ 
        Sag dSph  & MW &  $0.64 \pm 0.02$ & $11.4 \pm 0.7$  & $25.2 \pm 0.3$ & 9 & M12, I94\\
        And I  & A &  $0.22 \pm 0.04$ & $10.6 \pm 1.1$ & $24.7 \pm 0.2$ &  11& M12, M05, K10\\
        And III & A &  $0.52 \pm 0.02$ & $4.7 \pm 1.8$ & $24.8 \pm 0.2$ & 7&  M12, M05, K10\\
        And X & A &  $0.44 \pm 0.06$ &  $3.9 \pm 1.2$  & $26.3 \pm 1.1$ & 24& M12, K10\\
        And XIV  & A &  $0.31 \pm 0.09$ & $5.4 \pm 1.3$ & $27.2 \pm 0.6$ & 31& M12, K10\\
        And II  & A &  $0.20 \pm 0.08$ &  $7.3 \pm 0.8$ & $24.5 \pm 0.2$ & 4.9& M12, M05, K10\\
        And VII  & A &  $0.13 \pm 0.04$ &  $9.7 \pm 1.6$ & $23.2 \pm 0.2$ & 4.2& M12, M05, K10\\
        Cetus & I &   $0.33 \pm 0.06$ &  $17.0 \pm 2.0$ & $25.0 \pm 0.2$ & 46& M12, M05 \\
        Tucana & I &  $0.48 \pm 0.03$ & $15.8^{+4.1}_{-3.1} $  & $25.0 \pm 0.1$ & 73 & M12, F09 \\
    \hline
    \hline
    \end{tabular}
    \caption{The observed Local Group dwarf spheroidal galaxies used in this study, compiled from Refs.~\cite{2012AJ....144....4M}[M12], \cite{2018arXiv180606891M}[M18b] and associated sources (\cite{Simon_2011}[S11], \cite{Kirby_2013}[K13a], \cite{Willman_2011}[W11], \cite{2013ApJ...779..102K}[K13b], \cite{2007ApJ...670..313S}[S07], \cite{Koposov_2011}[K11], \cite{Walker:2008ax}[W09a], \cite{Walker_2009}[W09b], \cite{Koch:2007ye}[K07],  \cite{1994Natur.370..194I}[I94], \cite{2005MNRAS.356..979M}[M05], \cite{Kalirai_2010}[K10], \cite{2009AA...499..121F}[F09]) . The most relevant structural and kinematic parameters are listed here: the ellipticity $\epsilon$, central surface brightness $\Sigma_*$, the stellar velocity dispersion $\sigma_*$, the mass-to-light ratio $M/L_V$, and affiliation -- whether it is a satellite of the Milky Way [MW], Andromeda [A], or an isolated dwarf [I].  We note that reported parameters for MW satellites in both catalogs are in excellent agreement, and we will take the Plummer structural parameters reported in Ref.~\cite{2018arXiv180606891M} for consistency. }
    \label{tab:obsgals}
\end{table}

We note that several new ultra-faint candidates have been discovered from the Dark Energy Survey (\cite{Drlica-Wagner:2015ufc,0004-637X-805-2-130, 2015AJ....150..150F, 2014SPIE.9149E..0VD, 0004-637X-807-1-50}); however, most have not yet been classified as dwarf spheroidals, and only a limited amount have had spectral observations that successfully resolved a stellar velocity dispersion (see e.g. Refs. \cite{2015ApJ...811...62K,Simon:2015fdw, 2016ApJ...819...53W} for observational results).  Furthermore, these ultra-faint dwarfs have stellar masses as low as $\sim 100 M_\odot$, and may not be well represented by the simulations at hand. Therefore we will limit our analysis to the well-observed and comparatively massive dwarfs listed in Refs. \cite{2012AJ....144....4M,2018arXiv180606891M} and note that it may be easily extended to new dsphs when higher resolution data and simulations become available.

\begin{figure}[h!!]
    \centering
    \subfloat[Relationship between stellar mass $M_*$ and dynamical mass $M_{dyn}$ enclosed in half-light radius for simulated and observed LG dwarf galaxies]{
        \includegraphics[width=0.45\textwidth]{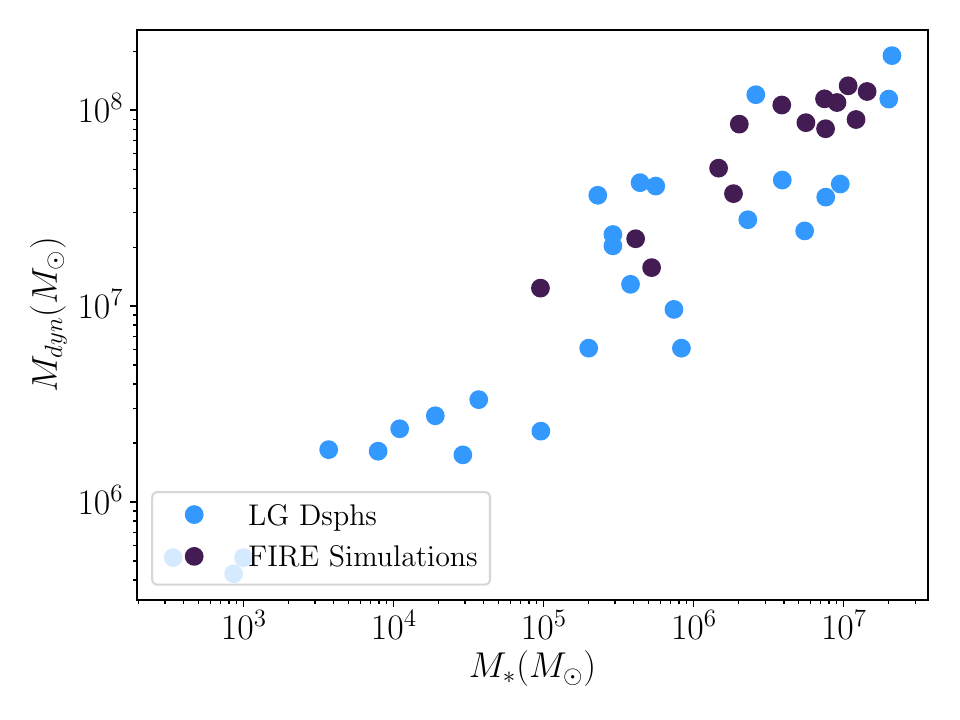}}
~
    \subfloat[Relationship between stellar mass $M_*$ and {3-D} half-light radius $r_{1/2}$ for simulated and observed LG dwarf galaxies]{
        \includegraphics[width=0.45\textwidth]{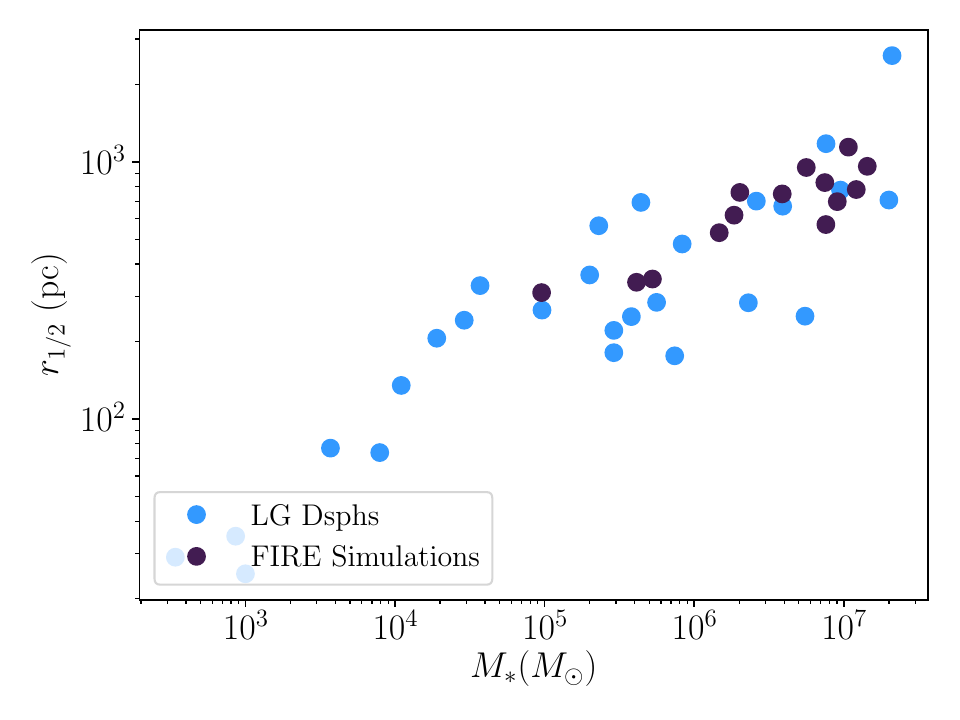}}
        
    \subfloat[Distribution of mass-to-light ratios for simulated and observed LG dwarf galaxies]{
        \includegraphics[width=0.45\textwidth]{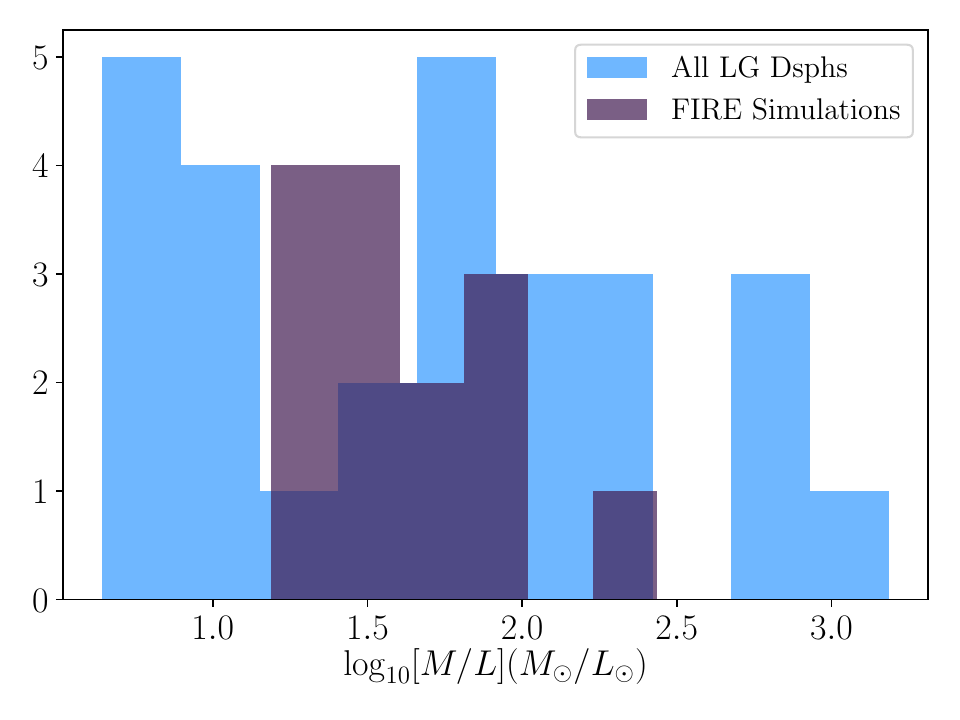}}
    \caption{Some structural properties of observed and simulation galaxies:  the distributions of $M_*$ in relation to $M_{dyn}$ within half-light radius and the {3-D} half-light radius $r_{1/2}$ itself, and the explicit distribution of mass-to-light ratios. The FIRE simulations provide a good representation of moderately massive {and moderately bright (mass-to-light $10 M_\odot/L_\odot < M/L< 100 M_\odot/L_\odot$)} dwarf galaxies, as they are resolution-limited in the low-mass regime.}\label{obsstats}
\end{figure}

To inform properties of dwarf galaxies explicitly sourced by CDM, we use the Feedback in Realistic Environments (FIRE) suite of cosmological simulations (\cite{Fitts:2016usl}). This set of hydro-intensive simulations contains 15 isolated dark matter halos, 14 of which exhibit star formation.  The halo masses are set to be of order $10^{10} M_\odot$ at $z=0$, and the stellar masses correspond roughly to the range of observed classical dwarf galaxies, $10^5-10^7 M_\odot$. The baryonic particle mass is set at $500 M_\odot$ and the dark matter particle mass is set at $2500 M_\odot$. Fig. \ref{obsstats} shows the relationship between stellar mass $M_*$, dynamical mass $M_{dyn}$, and {3-D} half-light radius $r_{1/2}$ for both the simulated and observed dwarf galaxies, as well as the explicit distribution of mass-to-light ratios $M_{dyn}/L_*$ for both, all assuming that stellar mass follows light.  As shown, FIRE galaxies appear in good agreement with moderately massive dwarf satellites, but we observe real galaxies at far lower masses, where the simulations are resolution-limited. {For the same reason, all but one of these simulated galaxies are relatively baryon-rich, that is their mass-to-light ratios are $< 100 M_{\odot}/L_\odot$, although they are also $> 10 M_{\odot}/L_\odot$.}  As resolution for hydrodynamic intensive simulations improves, the expectations for properties of low-mass CDM dwarfs will be better characterized, {and direct comparisons to analogous observed galaxies will be made possible.}

\begin{figure}[h!!]
    \centering
    \subfloat[Distribution of stellar axis ratios $b/a, c/a$ evaluated at half-light. As shown, the FIRE galaxies are largely prolate in stellar distribution]{
        \includegraphics[width=0.45\textwidth]{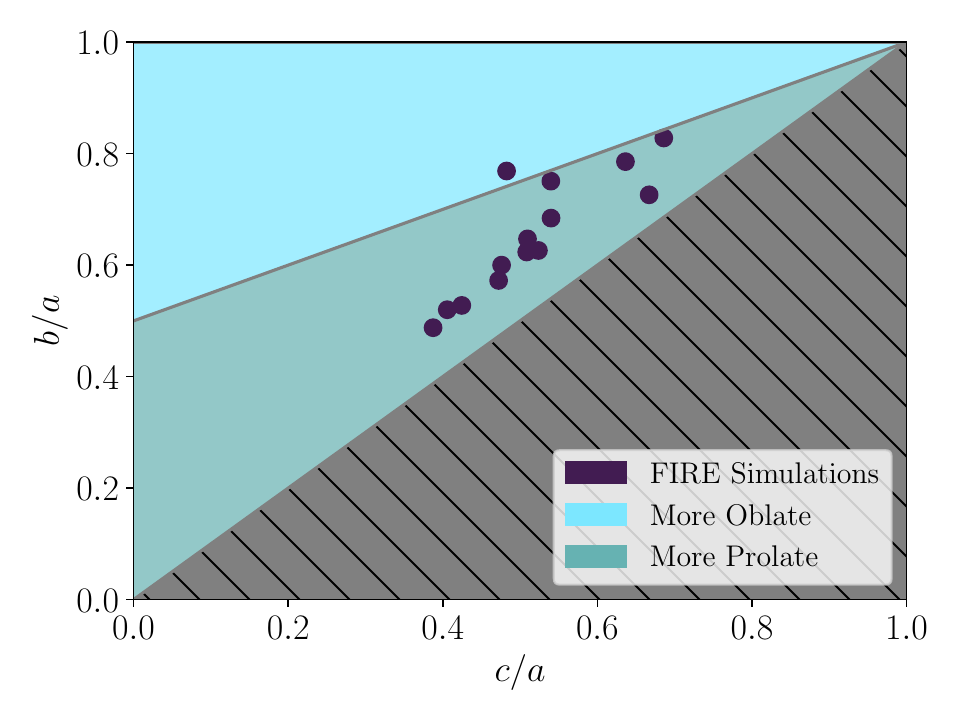}} 
~
    \subfloat[The half-light radius and stellar velocity dispersion measured along the semi-major (square) and semi-minor (circle) axes for each dwarf galaxy.]{
        \includegraphics[width=0.45\textwidth]{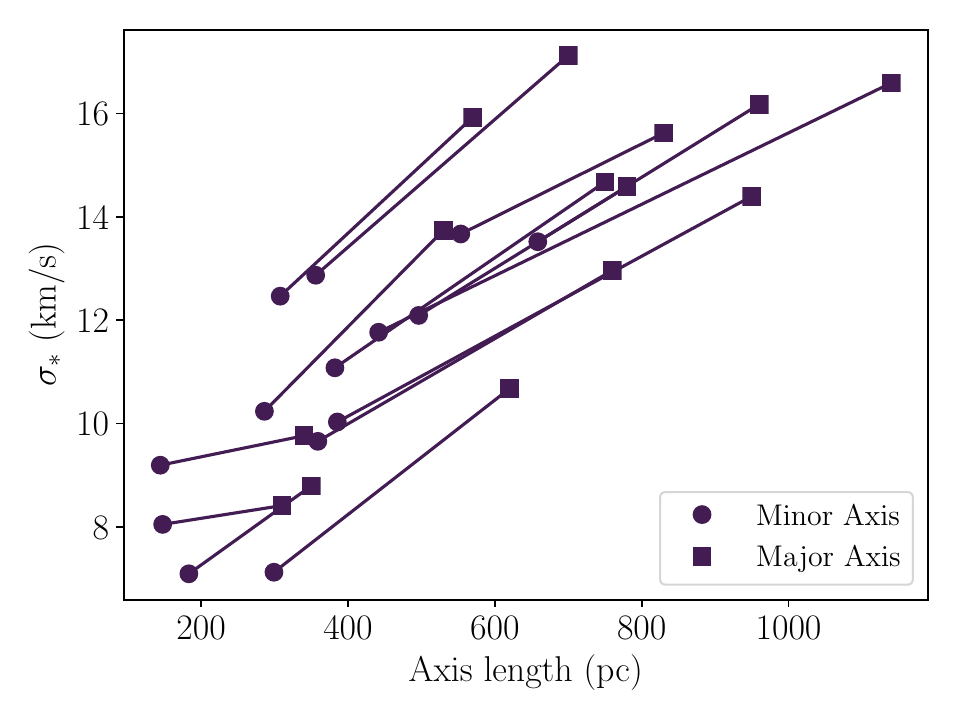}}
    \caption{Morphological properties of FIRE dwarf galaxies. As discussed in Section~\ref{GR}, if these galaxies were observed along random axes one expects anti-correlation between projected ellipticity and central surface brightness (stellar velocity dispersion). }\label{simstats}
\end{figure}

Fig. \ref{simstats} shows the distribution of stellar axis ratios of these simulated dwarfs, evaluated at half-light radius (left), and investigates the correlation between a galaxy's principal axis length and the magnitude of stellar velocity dispersion measured along that axis (right). The principal axes are computed from the galaxies' moments of inertia and the axis ratios are defined with respect to these. We find that the FIRE galaxies are indeed prolate in stellar distribution, and that the velocity dispersion measured along the semi-major axis is correspondingly larger than that measured along the semi-minor axis, and these set our expectations for dwarf spheroidal morphology under the CDM hypothesis.

\section{Up-Projection of Observation Data}\label{up}

Here we summarize results from deprojecting observed LG galaxies into a distribution of axis ratios $(b/a, c/a)$ using the formalism from the previous section. We assume a  uniform distribution of projection angles and separate the LG dwarfs fiducially into bright ($M/L < 100 M_\odot/L_\odot$)  and dim ($M/L > 100 M_\odot/L_\odot$)  samples. The division into subpopulations is motivated in section~\ref{down}.  For each observed dwarf we use only the observed ellipticity $\epsilon_{obs}$. The Bayesian likelihood of an underlying axis ratio $(b/a, c/a)$ given $\epsilon_{obs}$ is

 \begin{equation}P\left(\frac{b}{a}, \frac{c}{a} \mid \epsilon_{obs} \right) =    \frac{P(\epsilon_{obs} | \frac{b}{a}, \frac{c}{a}) P ( \frac{b}{a}, \frac{c}{a}) }{P(\epsilon_{obs})}  \end{equation}

To penalize against unphysical and extreme axis ratios, we adjust the prior $P(\frac{b}{a}, \frac{c}{a})$ to be proportional to the probability that this set of axis ratios projects into an ellipticity within the range of ellipticities that have been observed.

 \begin{equation} P \left(\frac{b}{a}, \frac{c}{a}\right) = \frac{1}{N}P(\epsilon_{proj} \in [\epsilon_{obs, min}, \epsilon_{obs, max}|) = \frac{1}{4\pi N}\int d\Omega \; \Theta(\epsilon_{proj} - \epsilon_{obs, min}) \Theta(\epsilon_{obs, max} - \epsilon_{proj}) \end{equation}

where $\epsilon_{proj}$ is a function of $b/a, c/a, \theta, \phi$ given in the previous section,  $\Theta$ denotes the Heaviside function, and $N$ is a constant for normalization. Here, we are sampling uniformly in $d\Omega = d\phi \; d\cos\theta$, but this can be easily adapted when the orientations of these galaxies relative to us are better known.

\begin{figure}[h!!]
    \centering
    \subfloat[Distribution of $b/a$ and $c/a$ for bright LG galaxies]{
        \includegraphics[width=0.45\textwidth]{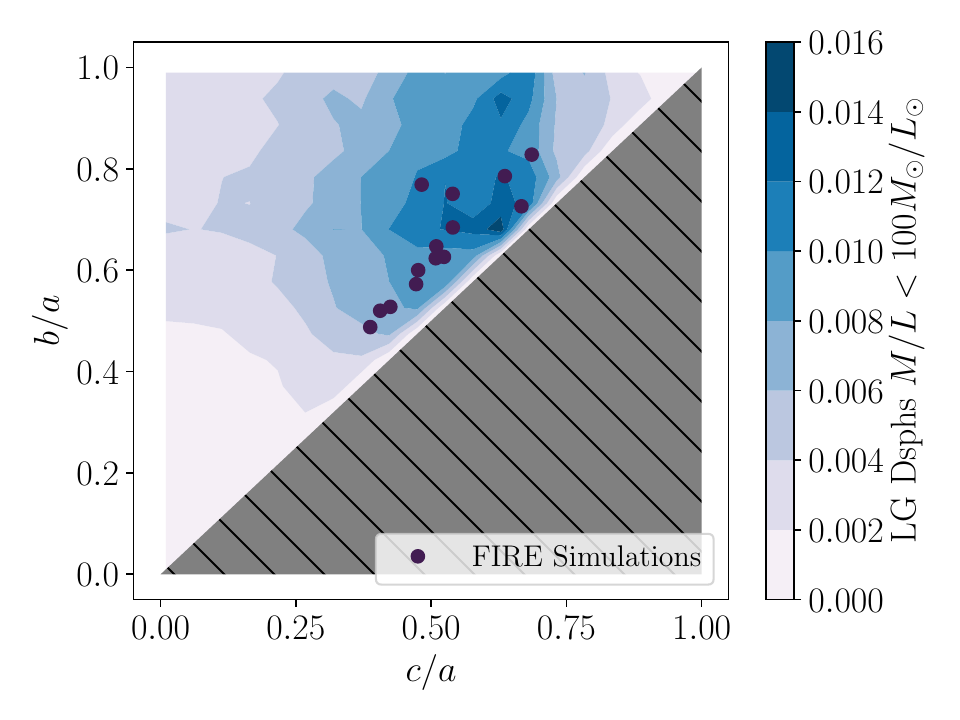}} 
    ~ 
    \subfloat[Distribution of $b/a$ and $c/a$ for dim LG galaxies]{
        \includegraphics[width=0.45\textwidth]{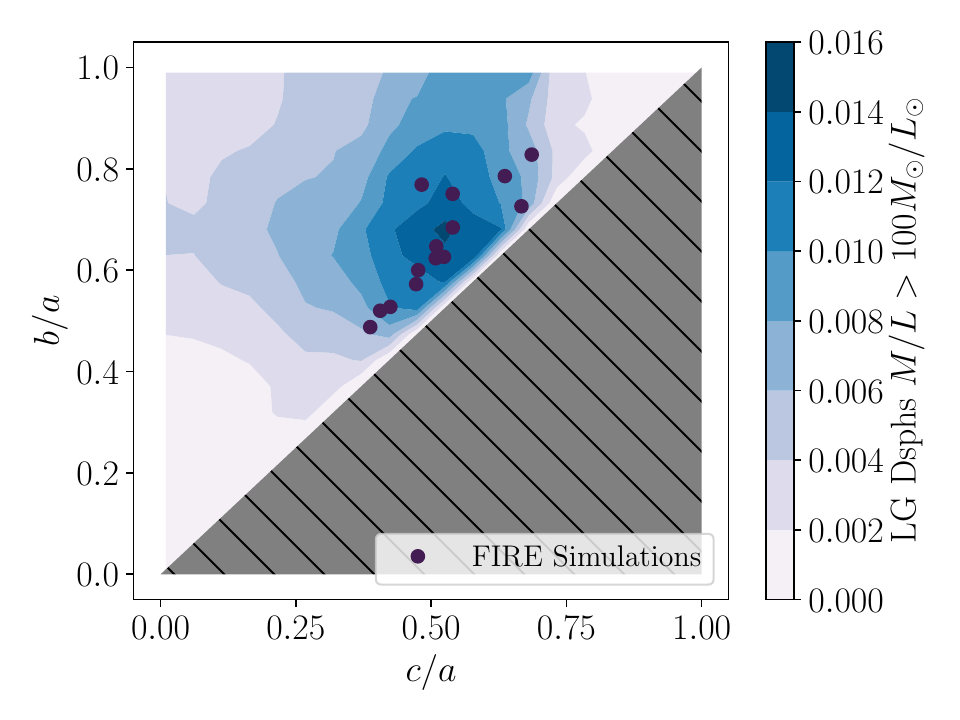}}
     
    \subfloat[Distribution of $b/a$ and $c/a$ for the full set of LG galaxies]{
        \includegraphics[width=0.45\textwidth]{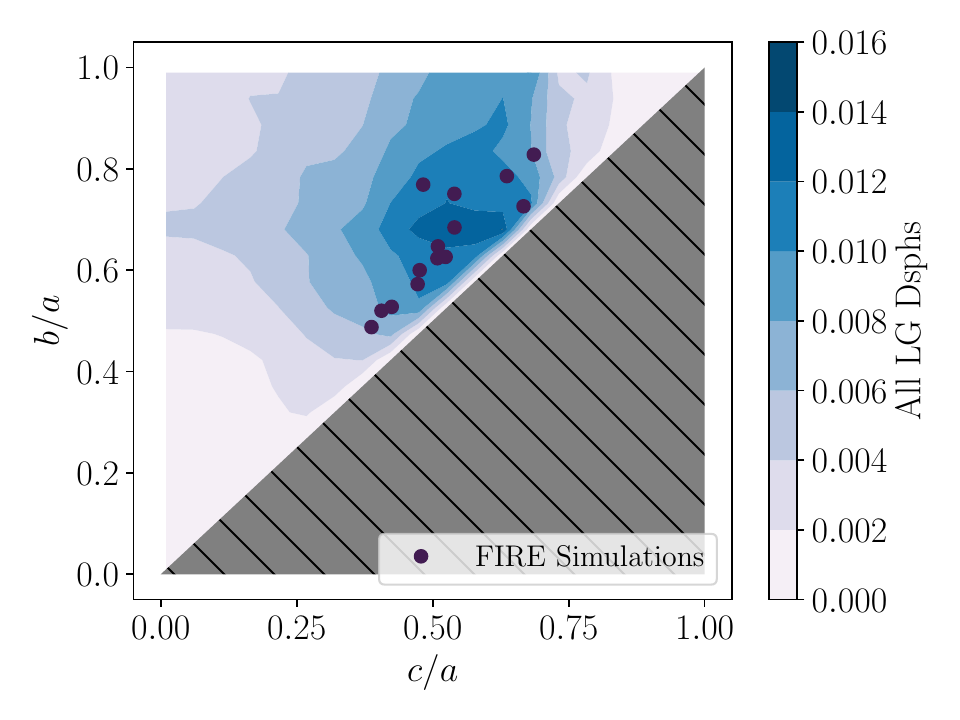}}
    \caption{Distribution of $b/a$ and $c/a$ of LG galaxies assuming uniform observing angles, compared to the FIRE simulated galaxies. The hatched region is forbidden by definition $b>c$; the diagonal $b=c$ corresponds to pure prolate shapes and the top edge $b/a=1$ corresponds to pure oblate shapes. The FIRE galaxies exhibit generally prolate axis ratios.}\label{axisrat}
\end{figure}

Fig.~\ref{axisrat} then gives the distribution of deprojected axis ratios of bright, dim, and the full-set of LG dwarfs as compared to the axis ratios of FIRE dwarf simulations.  As shown, while the dim LG dwarfs appear to prefer slightly more elliptical axis ratios as compared to the bright sample, there is no significant inconsistency with the axis ratios from the FIRE galaxies. This shows that ellipticity alone shows no significant tension between LG and simulated dwarf galaxies.  In the next sections we will study the fitness of CDM with observations when other observables are considered in tandem.

\section{Down-Projection of Simulation Data}\label{down}

In this section we summarize results of comparing CDM+Hydro simulation galaxies with observed dwarf spheroidals in the local group by projecting down the 3D simulated galaxies via a distribution of projection angles.  

In this work we focus on two specific points of comparison: the correlation of ellipticity $\epsilon$ with central surface brightness $\Sigma_*$, and that of ellipticity $\epsilon$ with stellar velocity dispersion $\sigma_*$.  These correlations are quantified by correlation coefficients, defined as
 \begin{equation} r_{\epsilon \Sigma_*} = \frac{\sum_i (\epsilon_i - \bar\epsilon)(\Sigma_{*,i} -  \bar \Sigma_* )}{\sqrt{\sum_i (\epsilon_i - \bar\epsilon)^2 \sum_i (\Sigma_{*,i} - \bar\Sigma_*)^2 }} \qquad r_{\epsilon \sigma_*} = \frac{\sum_i (\epsilon_i - \bar\epsilon)(\sigma_{*,i} -  \bar \sigma_*)}{\sqrt{\sum_i (\epsilon_i - \bar\epsilon)^2 \sum_i (\sigma_{*,i} - \bar\sigma_*)^2 }}  \end{equation}
where overlines denote arithmetic average. The error from observations is propagated to an uncertainty in the correlation coefficient 
\begin{equation}
\delta r_{\epsilon \Sigma_* (\sigma_*)} ^2 = \left(\frac{\partial r_{\epsilon \Sigma_* (\sigma_*)}}{\partial \epsilon}\right)^2 \delta \epsilon^2 + \left(\frac{\partial r_{\epsilon \Sigma_*(\sigma_*)}}{\partial \Sigma_* (\sigma_*)}\right)^2 \delta \Sigma_*^2 (\sigma_*^2) + \left(\frac{\partial r_{\epsilon \Sigma_*( \sigma_*)}}{\partial \epsilon}\frac{\partial r_{\epsilon \Sigma_*(\sigma_*)}}{\partial \Sigma_* (\sigma_*)}\right)^2 \sigma_{\epsilon \Sigma_* (\sigma_*)}  
\end{equation} 
where $\sigma_{\epsilon \Sigma_* (\sigma_*)}$ denotes the sample covariance. As explained in Section~\ref{GR}, statistically significiant correlations or anti-correlations between ellipticity and central surface brightness (stellar velocity dispersion) should serve as useful proxies for the galaxy's morphology; a strong positive value in both $r_{\epsilon \Sigma_*}$ and $r_{\epsilon \sigma_*}$ is indicative of a collection of oblate galaxies, negative values are indicative of prolate bodies, and fully triaxial or mixed populations should show no correlation at all.

\begin{figure}[h!!]
    \centering
    \subfloat[Correlation of ellipticity $\epsilon = 1-b/a$ with central surface brightness $\Sigma_*$]{
        \includegraphics[width=0.45\textwidth]{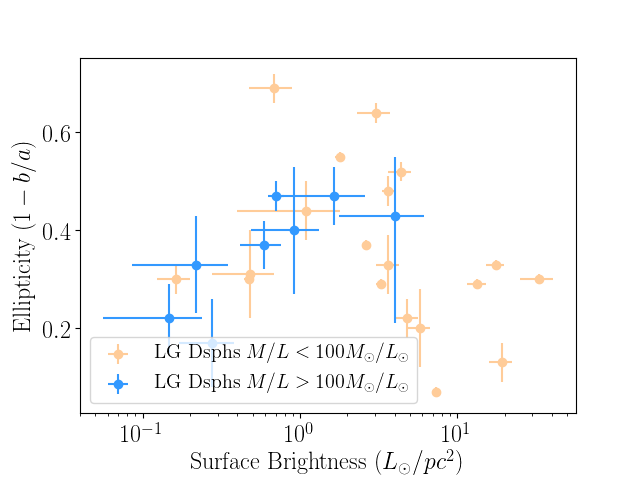}
        \label{fig:esb}} 
    ~ 
    \subfloat[Correlation of ellipticity $\epsilon = 1-b/a$ with stellar velocity dispersion $\sigma_*$]{
        \includegraphics[width=0.45\textwidth]{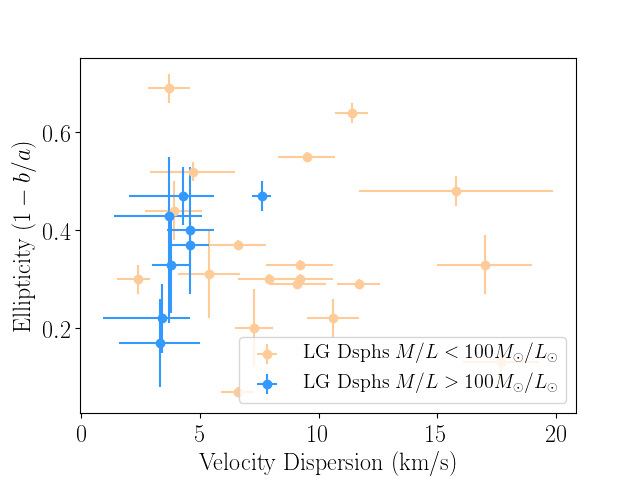}
        \label{fig:evd}}
    \caption{ Correlation of ellipticity and central surface brightness (left) and stellar velocity dispersion (right) of LG observed galaxies. Bright galaxies (fiducially $M/L  <100 M_\odot/L_\odot$) exhibit a generally negative correlation while dim galaxies ( $M/L  >100 M_\odot/L_\odot$) exhibit a generally positive one. }\label{eps-sigs}
\end{figure}

Fig~\ref{eps-sigs} shows the correlation of ellipticity and central surface brightness (left) and stellar velocity dispersion (right) of LG observed galaxies.  We note that for both observables there appears to be a separation between the correlation trends of brighter and dimmer galaxies -- that is, brighter galaxies appear to be negatively correlated in both, and the opposite is seen in dimmer galaxies. We split these two populations roughly into those with  $M/L  <100 M_\odot/L_\odot$ (``bright") and those with  $M/L  >100 M_\odot/L_\odot$ (``dim").  Specifically,

\begin{align}
r^{dim}_{\epsilon\Sigma_*} = 0.509 \pm 0.309 \qquad  r^{dim}_{\epsilon\sigma_*} =  0.618 \pm 0.323  \nonumber \\
r^{bright}_{\epsilon\Sigma_*} = -0.353 \pm 0.059  \qquad  r^{bright}_{\epsilon\sigma_*} = -0.183 \pm 0.098  \nonumber \\
r^{all}_{\epsilon\Sigma_*} = -0.287 \pm 0.058  \qquad  r^{all}_{\epsilon\sigma_*} = -0.105 \pm 0.097  \nonumber \\
\end{align}

\begin{figure}[h!!]
    \centering
    \subfloat[Distribution of $r_{\epsilon \Sigma_*}$ for FIRE dwarfs]{
        \includegraphics[width=0.45\textwidth]{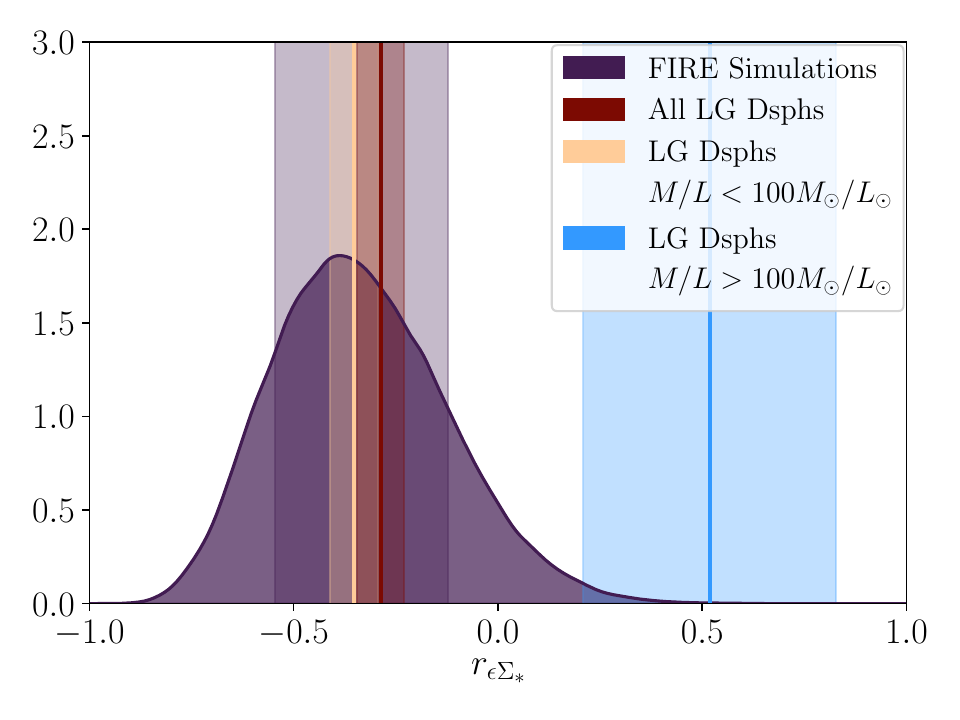}} 
    ~ 
    \subfloat[Distribution of $r_{\epsilon \sigma_*}$ for FIRE dwarfs]{
        \includegraphics[width=0.45\textwidth]{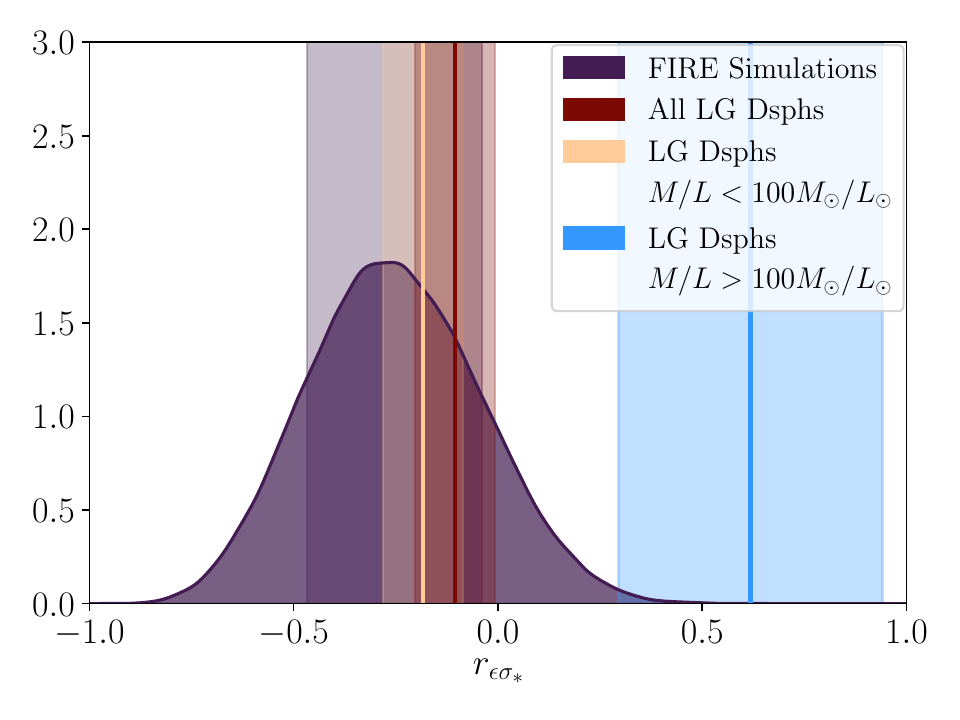}}
    \caption{ Distribution of correlation between ellipticity and central surface brightness (left) and stellar velocity dispersion (right) of uniformly projected FIRE galaxies. The measured correlations of bright, dim, and the full sample of LG dsphs are also shown. The shaded regions are $1\sigma$ limits. As shown, the population of galaxies with $M/L  >100 M_\odot/L_\odot$ is highly inconsistent with the simulated population.}\label{projectdown}
\end{figure}

\begin{figure}[h!!]
    \centering
        \includegraphics[width=\textwidth]{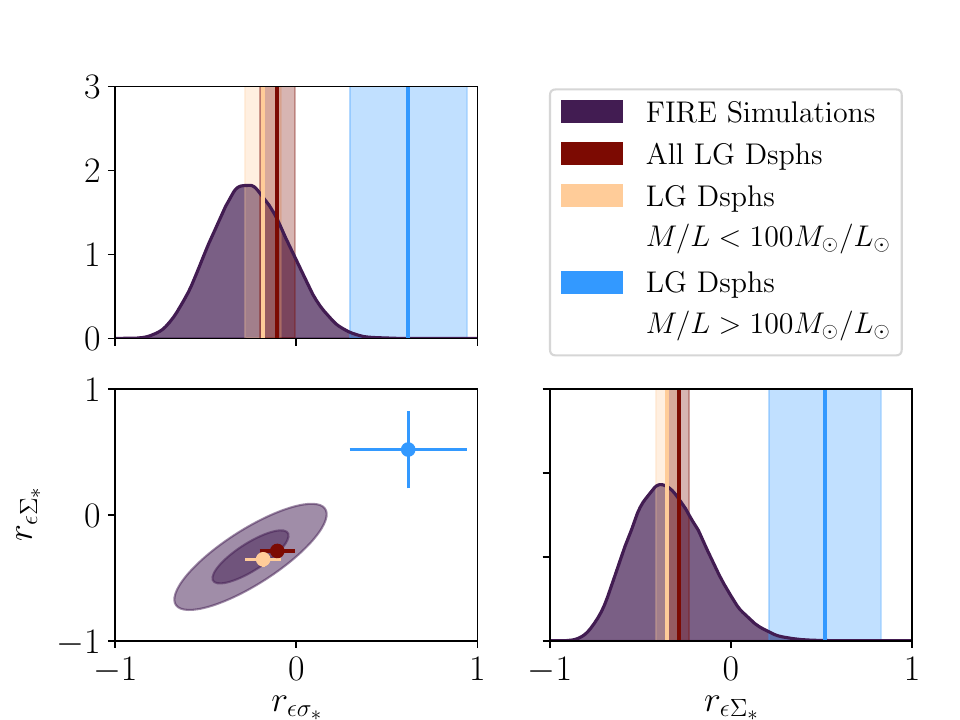}
    \caption{Similar to Fig.~\ref{projectdown} as a 2D-contour plot.} 
      \label{contour}
\end{figure}

On the simulation side, we project each of 14 FIRE dwarf galaxies along a random orientation and compute the $\epsilon-\Sigma_*$ and $\epsilon-\sigma_*$ correlation coefficients of the entire suite of galaxies. This process is repeated with uniformly random projection angles $10^5$ times, and distributions of $r_{\epsilon\Sigma_*}$ and $r_{\epsilon\sigma_*}$ are constructed.  These two correlation coefficients then serve as a test for morphological consistency between simulated and observed galaxies.  We obtain for the FIRE galaxies, assuming a uniform distribution of observation angles, 

\begin{equation}
r^{FIRE}_{\epsilon\Sigma_*} = -0.322\pm 0.213  \qquad  r^{FIRE}_{\epsilon\sigma_*} = -0.252\pm 0.214 
\end{equation}

which is marginally consistent with 0 in $\epsilon-\Sigma_*$ but with negative expectation in both, reflecting the overall prolate nature of simulated $\Lambda$CDM galaxies.

Fig~\ref{projectdown} shows the distributions of $r_{\epsilon  \Sigma_*}$ and  $r_{\epsilon  \sigma_*}$ of randomly projected FIRE galaxies. The measured correlations of bright, dim, and the full set of LG dwarfs are also shown. Fig~\ref{contour} shows where these correlations lie in $r_{\epsilon  \Sigma_*} - r_{\epsilon\sigma_*}$ space. {We reiterate that the FIRE Simulations consist of isolated and thus predominantly bright dwarfs, and are thus most directly counterparts of the bright LG dwarfs, but it is interesting still that there appears to be distinct morphological separation between the bright and dim subpopulations. Indeed as shown, while there is good agreement between the FIRE simulations and the bright galaxies,} the dim galaxies lie $\gtrsim 3\sigma$ away from the simulated galaxy expectation in both $r_{\epsilon \Sigma_*}$ and $r_{\epsilon \sigma_*}$, and furthermore lie $\gtrsim 4\sigma$ away from simulated galaxies in $r_{\epsilon \Sigma_*} - r_{\epsilon\sigma_*}$ space.  It is interesting to note that the disparity is much larger in ellipticity-central surface brightness correlation than in ellipticity-stellar velocity dispersion correlation, which is consistent with the observed and simulational understanding of dwarf spheroidals being isothermal to first order (\cite{ 2015ApJ...808..158B, annurev.astro.36.1.435}).  In this case, with the absence of rotational support, velocity-dispersion anisotropies may be attributed to anisotropies of the gravitational potential.

\section{Statistical Tests}\label{tests}

Thus far, we have presented evidence that the dim population of observed dsphs exhibit correlation between ellipticity and central surface brightness (stellar velocity dispersion) consistent with an oblate morphology, and both the bright population and FIRE simulations exhibit anti-correlation consistent with a prolate one. This is our main result. However, due to the limited size of both available simulational and observational dwarf galaxies, it is important to evaluate the rigor of this statement with care.  In particular, since the number of dim galaxies with $M/L > 100 M_\odot/L_\odot$ is small (eight in total), on the outset it is unclear that the positive correlation between ellipticity and central surface brightness (stellar velocity dispersion) is not an accident.  

The measurement uncertainties of the observables and resolution limitations of the simulations will be improved upon by future efforts, but in this section we address three additional sources of uncertainty:  the statistical uncertainty that the finite sample of simulations and observations are good estimators for the underlying distributions, the placement of the bright/dim population divisor, and the unknown true distribution of projection angles.

\subsection{Subsampling}

We investigate first whether the positive correlation seen in the dim dsphs can be obtained by chance via random sampling of  the bright subpopulation or the full set of LG galaxies.

\begin{figure}[h!!]
    \centering
    \subfloat[Distribution of $r_{\epsilon \Sigma_*}$ for subsampled full set and bright LG dwarfs]{
        \includegraphics[width=0.45\textwidth]{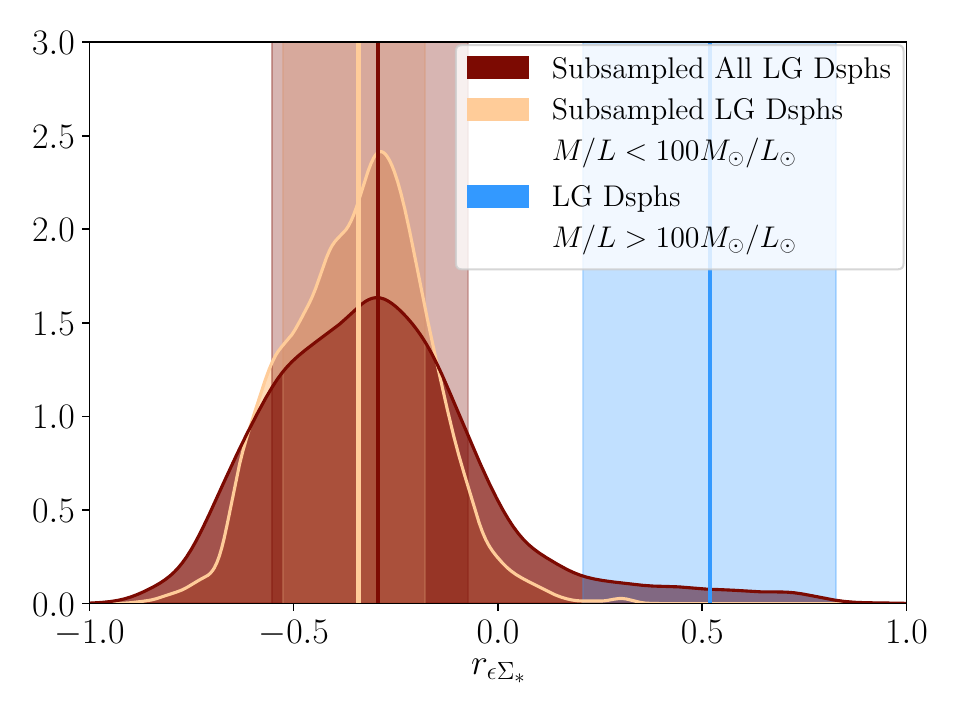}} 
~
    \subfloat[Distribution of $r_{\epsilon \sigma_*}$ for subsampled full set and bright LG dwarfs]{
        \includegraphics[width=0.45\textwidth]{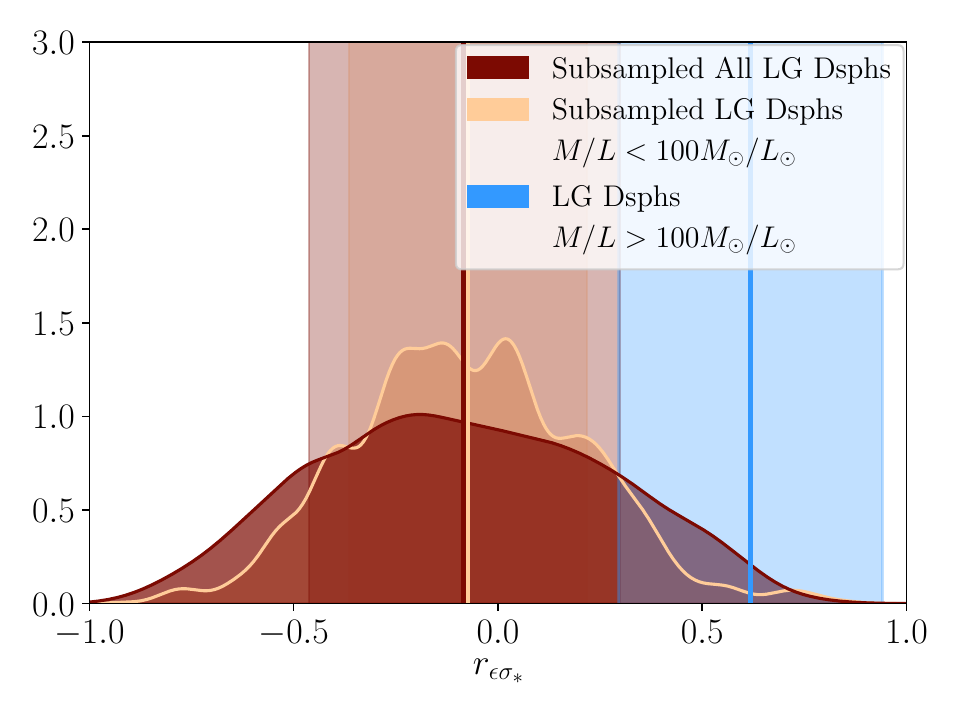}}
    \caption{ Distribution of correlation between ellipticity and central surface brightness (left) and stellar velocity dispersion (right) of a random subsample of 8 LG observed galaxies and LG Bright galaxies (fiducially $M/L  <100 M_\odot/L_\odot$).  The $1\sigma$ limits are shown as shaded regions. As shown, the positive correlation depicted by  dim galaxies ( $M/L  >100 M_\odot/L_\odot$) is not obtainable by random subsampling of the full LG population.}\label{subsamp1d}
\end{figure}

\begin{figure}[h!!]
    \centering
        \includegraphics[width=\textwidth]{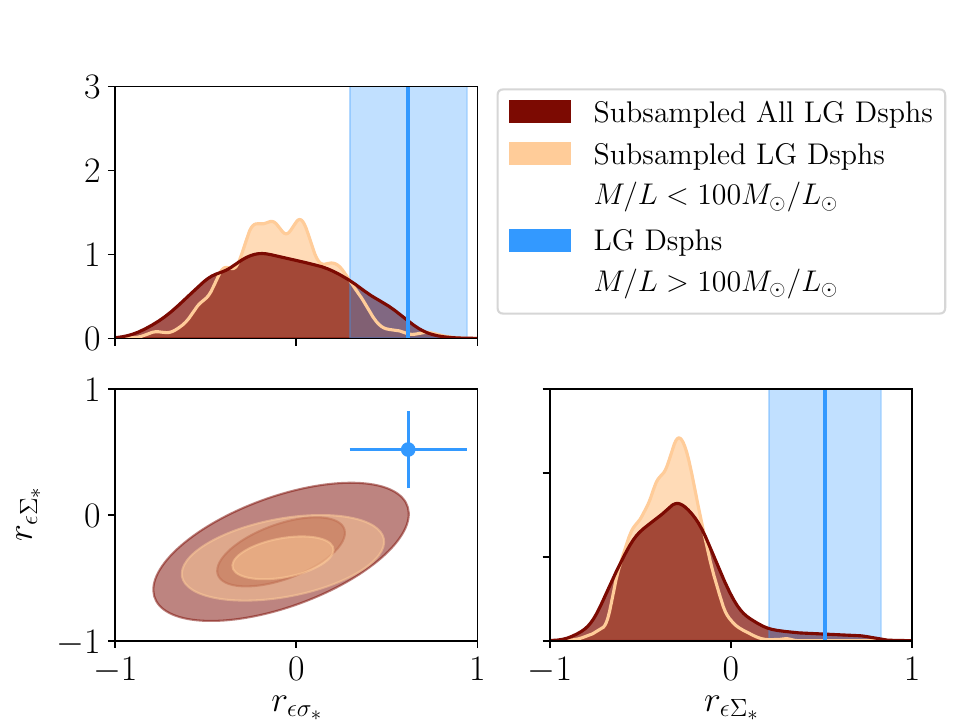}
    \caption{Similar to Fig.~\ref{subsamp1d} as a 2D-contour plot. } 
      \label{subsamp_contour}
\end{figure}

Fig.~\ref{subsamp1d} shows the distributions of correlation coefficients $r_{\epsilon \sigma_*}$ and $r_{\epsilon \Sigma_*}$ upon repeated subsampling of eight random galaxies within the full LG population (red) and the subpopulation of bright galaxies (orange).  Fig ~\ref{subsamp_contour} shows the corresponding 2D-distributions. We show that a random subsampling of currently observed LG dsphs cannot recover the correlation between ellipticity and central surface brightness (stellar velocity dispersion) that is seen in the dim subpopulation at the $\sim2\sigma$ level.  

\begin{figure}[h!!]
    \centering
    \subfloat[Distribution of $r_{\epsilon \Sigma_*}$ for subsampled FIRE galaxies]{
        \includegraphics[width=0.45\textwidth]{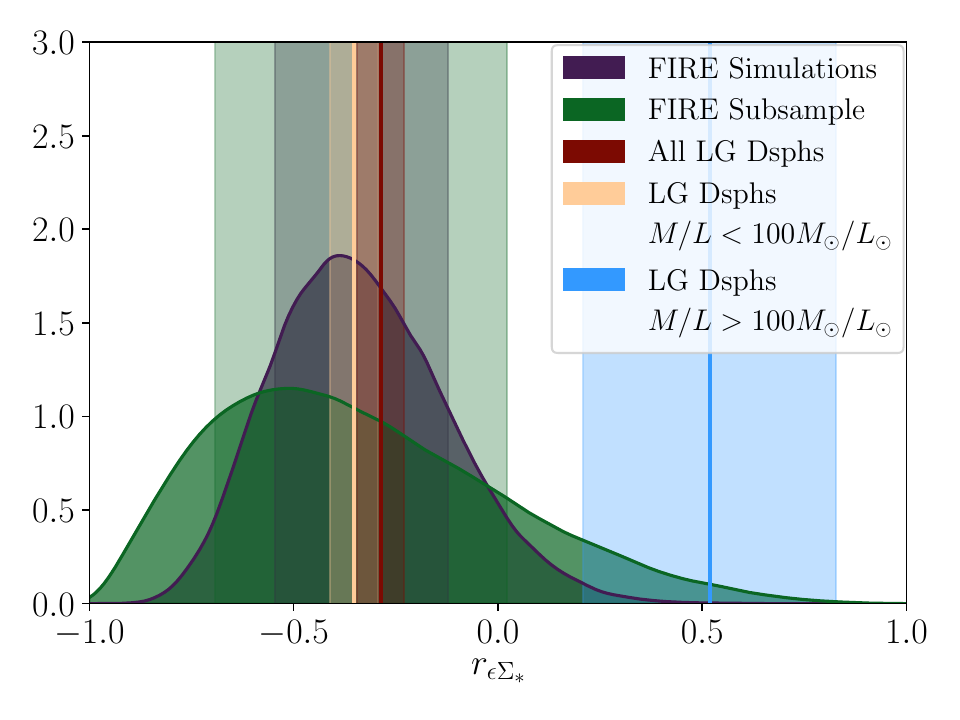} }
~
    \subfloat[Distribution of $r_{\epsilon \sigma_*}$ for subsampled FIRE galaxies]{
        \includegraphics[width=0.45\textwidth]{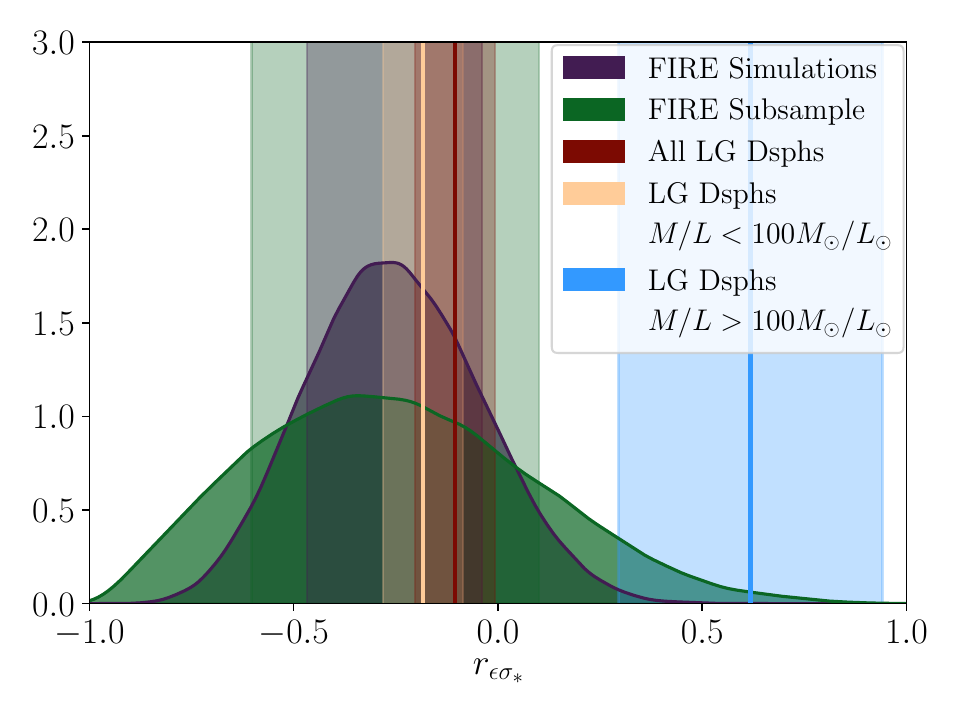}}
    \caption{ Distribution of correlation between ellipticity and central surface brightness (left) and stellar velocity dispersion (right)  of a random subsample of 8 FIRE simulated galaxies, compared with LG observed galaxies and the distribution from the full FIRE galaxy set..  The $1\sigma$ limits are shown as shaded regions. As shown, subsampling significantly widens the distribution of both coefficients and the bright and full set of LG dwarfs become consistent with the FIRE subsample, but the dim LG dwarfs do not.}\label{firesubsamp1d}
\end{figure}

\begin{figure}[h!!]
    \centering
        \includegraphics[width=\textwidth]{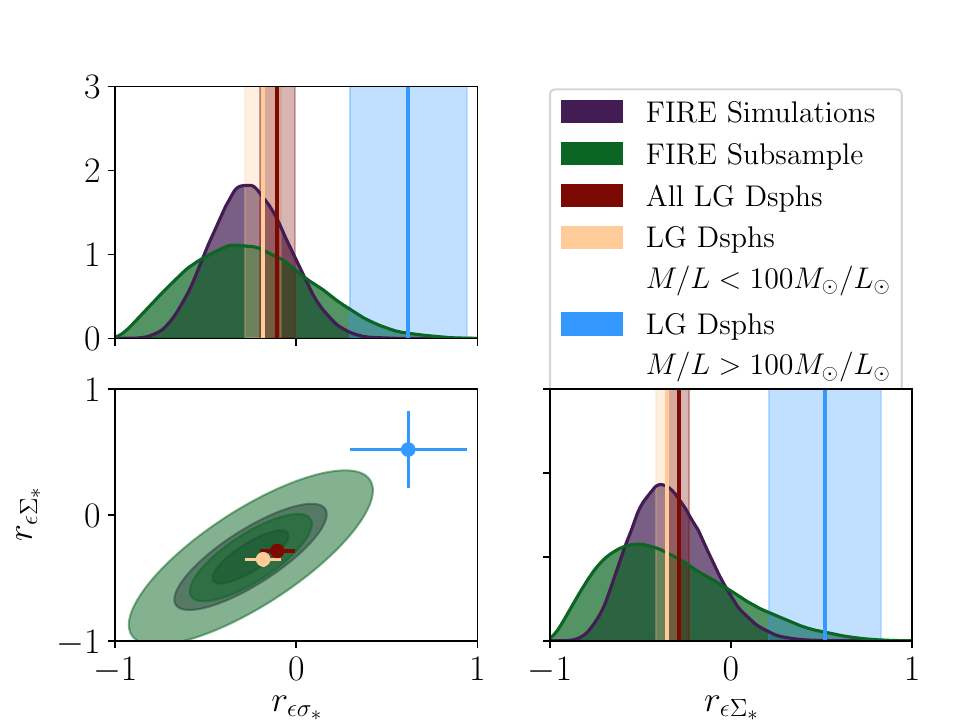}
    \caption{Similar to Fig.~\ref{firesubsamp1d} as a 2D-contour plot.} 
      \label{firesubsamp_contour}
\end{figure}

We study also the broadening of the correlation coefficient distribution for FIRE galaxies via subsampling eight random galaxies and randomly projecting each one.  As shown in Figs.~\ref{firesubsamp1d} and ~\ref{firesubsamp_contour}, the distributions widen significantly upon taking a subsample of roughly half the full size, as expected. However, the dim LG galaxies still retain a $>3\sigma$ tension.

Thus it appears that neither the discrepancy between dim LG dsphs and the simulation dwarfs nor the discrepancy between dim LG dsphs and their brighter counterparts may be satisfactorily explained by subsampling -- it is extremely unlikely that a random choice of eight galaxies from either the data or simulations could have produced the very positive correlations exhibited by the eight dim galaxies. It is more likely that this subpopulation is genuinely morphologically distinct.

\subsection{Choice of Bright-Dim Threshold}

\begin{figure}[h!!]
    \centering
    \subfloat{
        \includegraphics[width=0.45\textwidth]{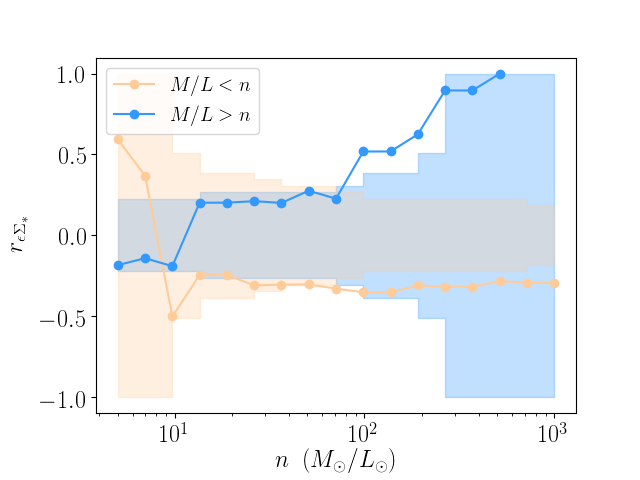}}
~
    \subfloat{
        \includegraphics[width=0.45\textwidth]{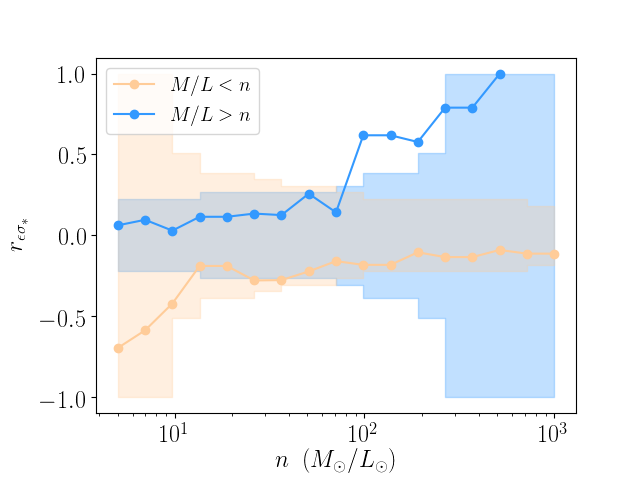}}
    \caption{The ellipticity-central surface brightness (left) and ellipticity-stellar velocity dispersion (right) correlation of bright (orange) and dim (blue) dwarf galaxies for various choices $n$ as the division between ``bright" and ``dim".  A choice of $n$ at too low or too high a mass-to-light ratio is unideal since it restricts the size of one of the populations. The shaded regions show the regions in correlation space for both bright and dim populations that is consistent with an underlying $r=0$ at the $68\%$CL for each choice of bright-dim threshold. A cut at $\sim 100 M_\odot/L_\odot$ is a good position to isolate the correlated component, while not excluding too many galaxies such that small number uncertainties dominate the correlated result.}\label{cuts}
\end{figure}

In Fig.~\ref{cuts} we investigate the sensitivity of our results to the placement of this bright-dim division point.   We find that  a cut at $O( 100) M_\odot/L_\odot$ is necessary to observe a difference in correlation of ellipticity-stellar velocity dispersion, but correlation between ellipticity-central surface brightness and the separation between bright and dim populations is not very sensitive to the precise choice of division point.  We also observe that the correlation at the high mass-to-light, or dim,  end of the sample is quite stable for both sets of observables.  

The position of the bright-dim threshold also controls the number of galaxies in each population, and for a sparser population a stronger correlation or anti-correlation must be observed for the result to be significant.  The expected distribution of correlation coefficients measured from taking $N$ samples from a underlying distribution that is intrinsically uncorrelated is given by 

   \begin{equation}p(r) =  \frac{(1-r^2)^{\frac{N-4}{2}}}{B(\frac{1}{2}, \frac{N-2}{2})} \end{equation}

where $B (x,y)$ is the Beta function. In Fig~\ref{cuts} we show also the region in correlation space for both bright and dim populations that is consistent with an underlying $r=0$ at $1\sigma$ for each choice of bright-dim threshold.  As shown, the separation between dim and bright subpopulations of LG dwarfs, and between both populations and the expected range is consistent with no correlations, is most enhanced for a cut at $70-200 M_\odot/L\odot$. However, this separation is still marginal and future measurements of additional galaxies are necessary to confirm the proposed separation between bright and dim dwarf galaxies.

\subsection{Distribution angles}

\begin{figure}[h!!]
    \centering
        \subfloat{
        \includegraphics[width=0.45\textwidth]{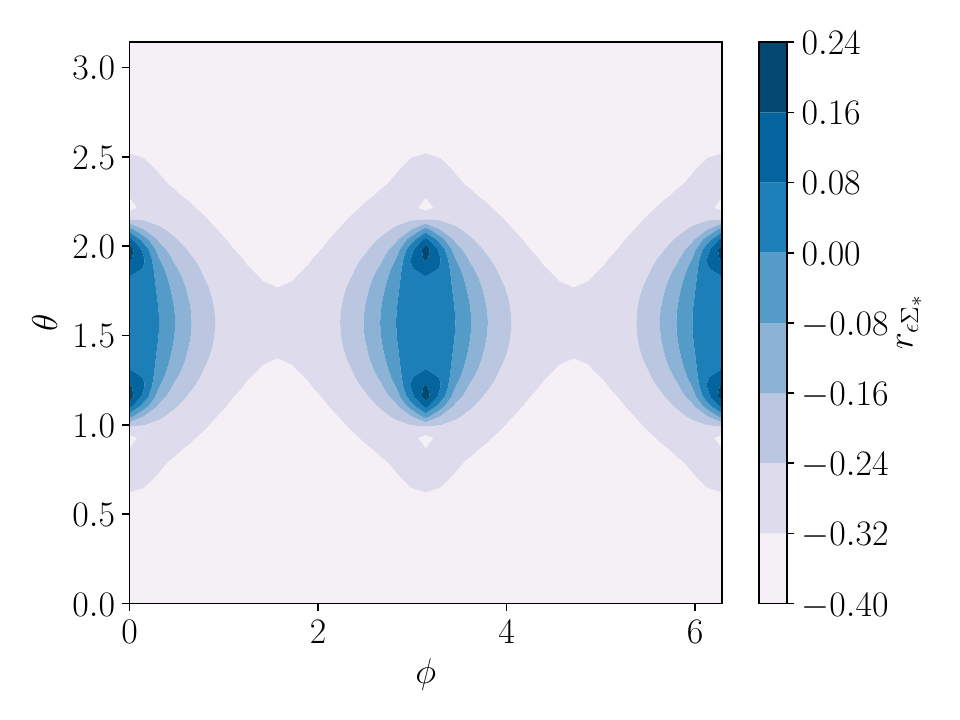}}
    ~
    \subfloat{
        \includegraphics[width=0.45\textwidth]{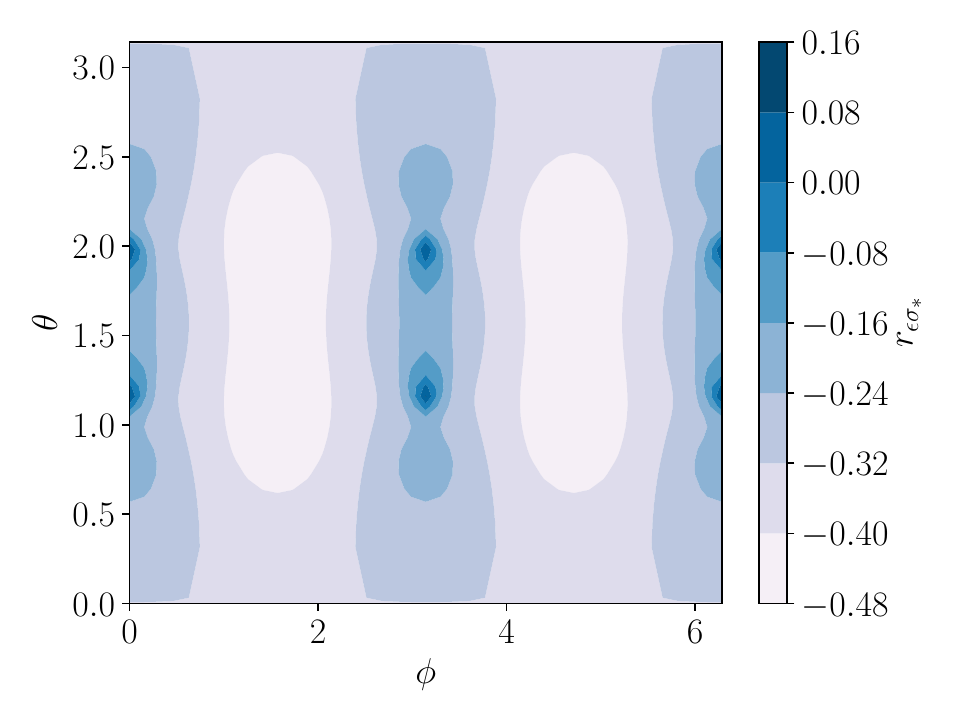}}
    \caption{Induced distribution of $r_{\epsilon \Sigma_*}$ and $r_{\epsilon \sigma_*}$ from observing all FIRE galaxies at angle $(\theta, \phi)$}\label{angles}
\end{figure}

In this section we investigate the effect of relaxing the assumption of uniform observation angle distributions. In practice the distribution of observation angles is non-uniform, as satellite galaxies experience tidal locking with their host and tidal disruption with each other.  Fig.\ref{angles} shows the induced correlation coefficients upon viewing all FIRE galaxies at each angle $(\phi, \theta)$, to check if a single preferred viewing angle may induce the positive correlation seen in the dim LG dsphs sample.  As shown, the maximum amount of correlation in $r_{\epsilon \Sigma_*}$ obtainable in this manner is $\sim 0.3$ and  the maximum obtainable in  $r_{\epsilon \sigma_*}$ is $\sim 0.1$, far below what is needed to explain the $\gtrsim 0.7$ correlation in the dim dsphs.

\begin{figure}[h!!]
    \centering
        \subfloat{
        \includegraphics[width=0.3\textwidth]{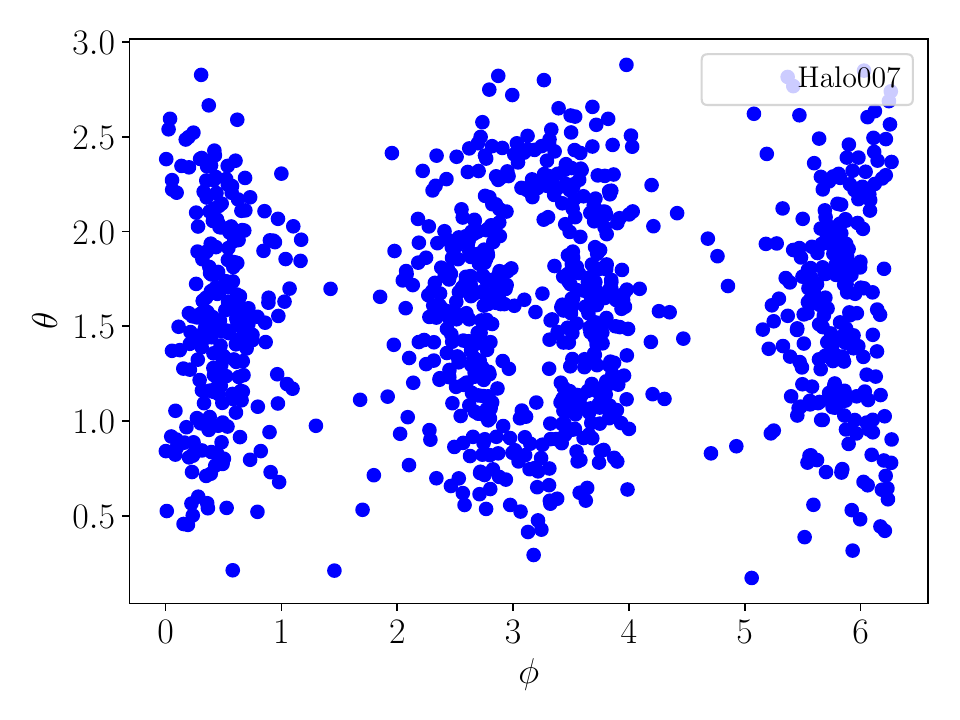}}
    ~ 
    \subfloat{
        \includegraphics[width=0.3\textwidth]{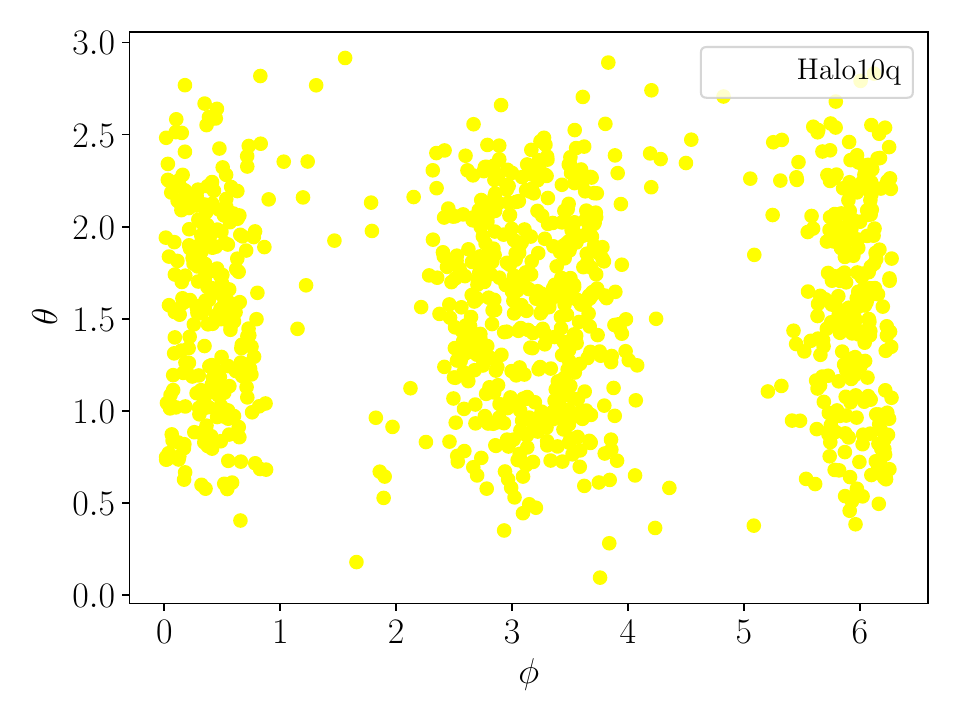}} 
~
  \subfloat{
        \includegraphics[width=0.3\textwidth]{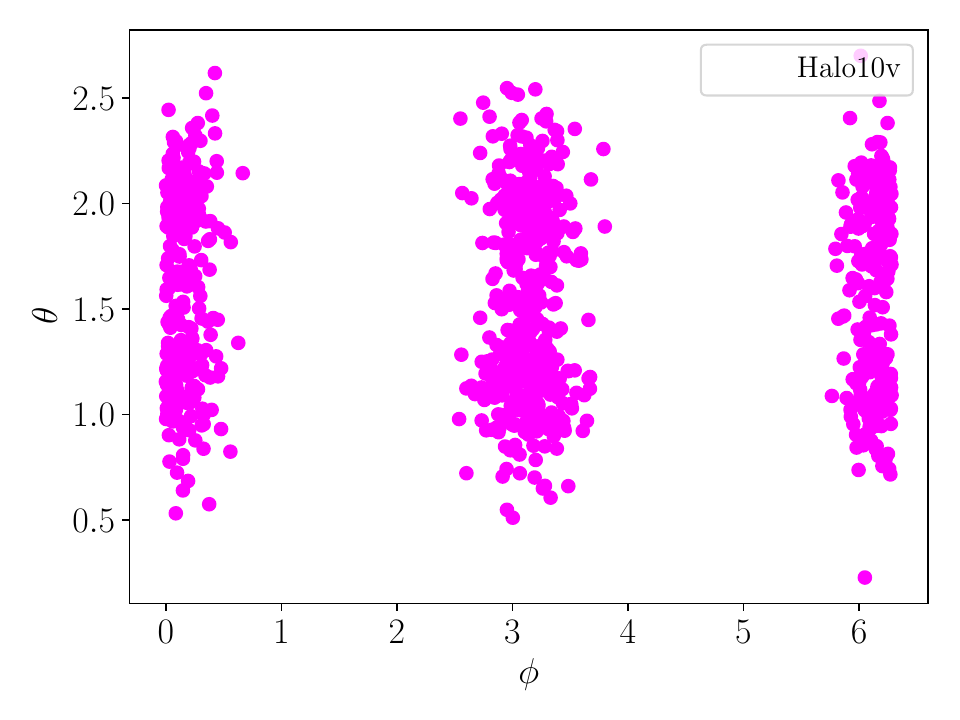}}\\
    \subfloat{
        \includegraphics[width=0.3\textwidth]{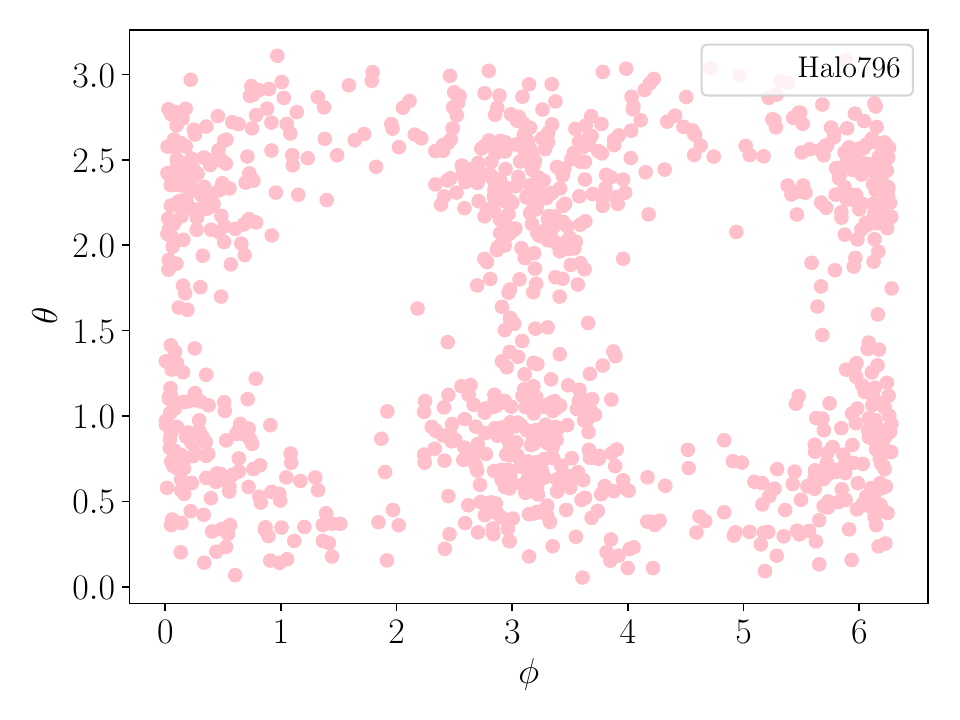}}
  \subfloat{
        \includegraphics[width=0.3\textwidth]{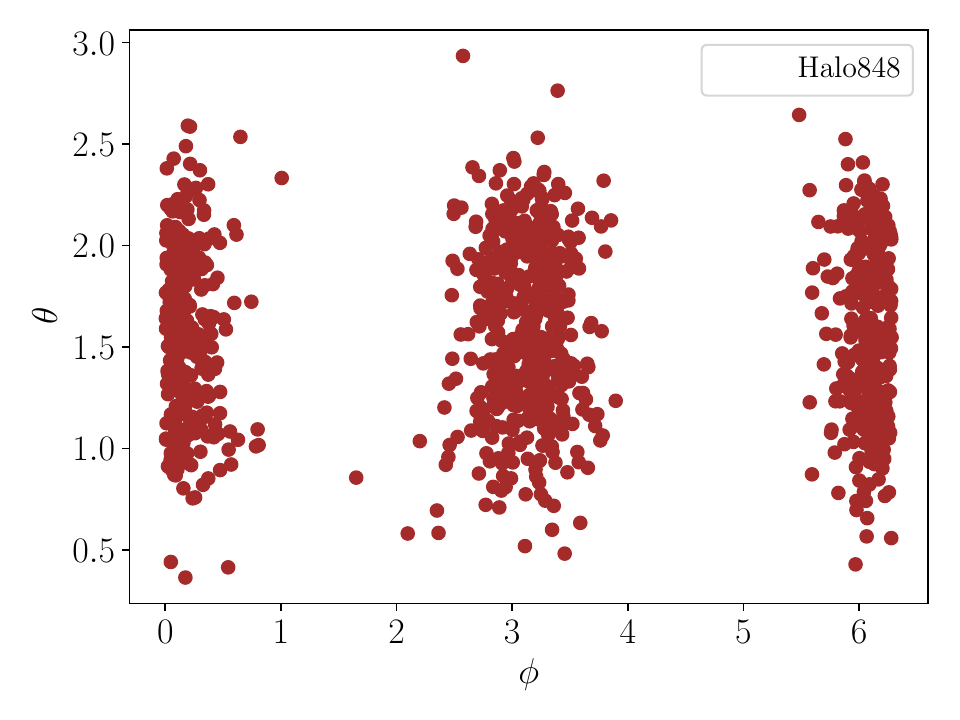}}
    ~ 
    \subfloat{
        \includegraphics[width=0.3\textwidth]{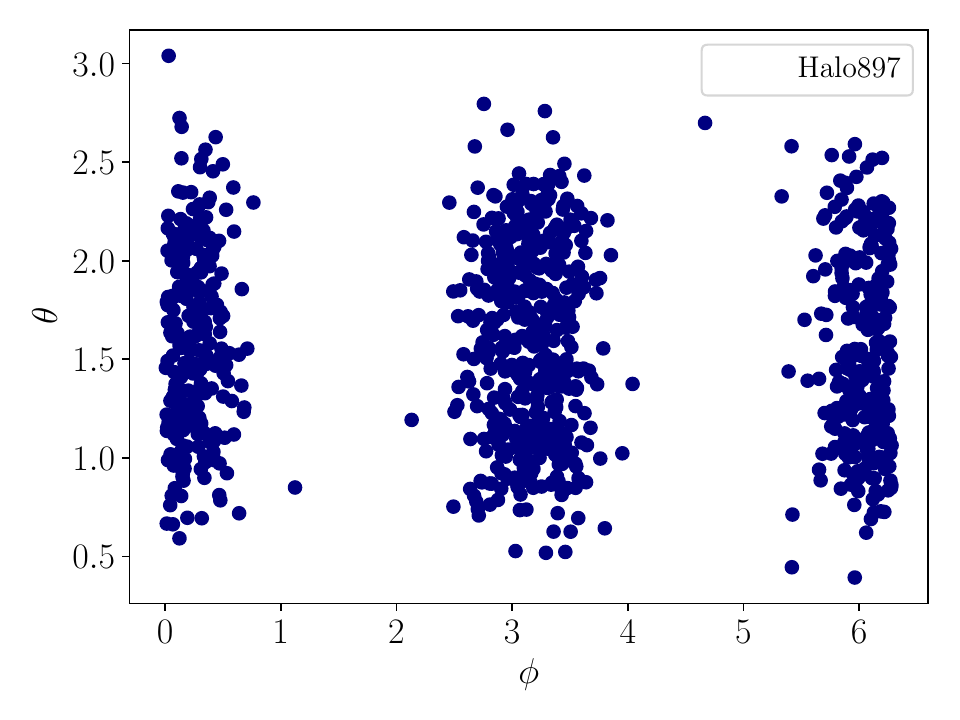}}\\
  \subfloat{
        \includegraphics[width=0.3\textwidth]{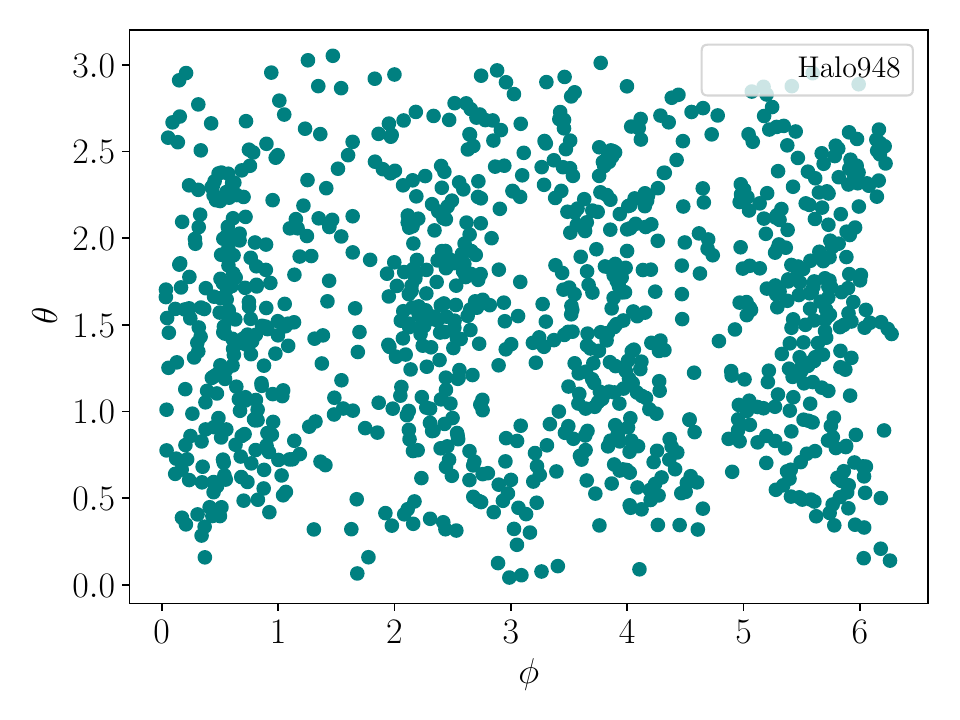}}
    ~ 
    \subfloat{
        \includegraphics[width=0.3\textwidth]{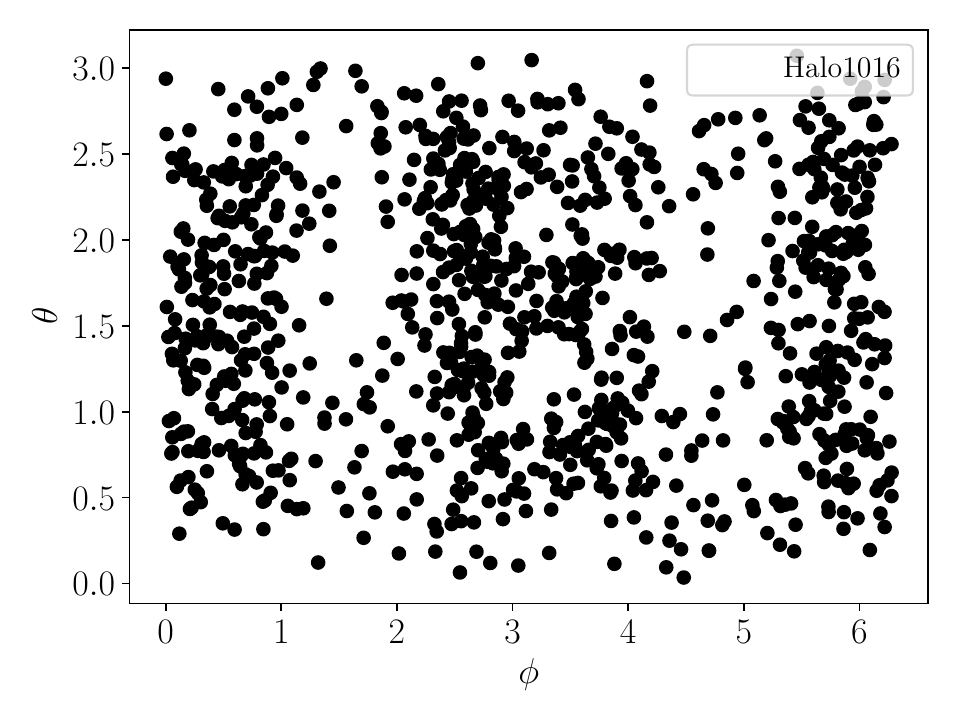}}
  \subfloat{
        \includegraphics[width=0.3\textwidth]{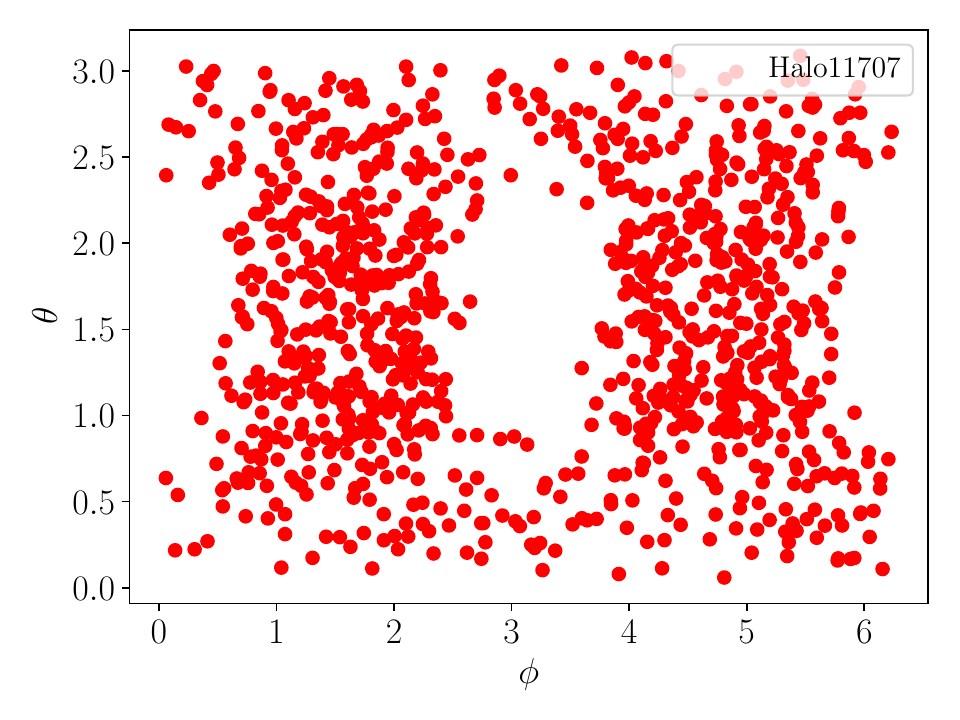}}\\
    ~ 
    \subfloat{
        \includegraphics[width=0.3\textwidth]{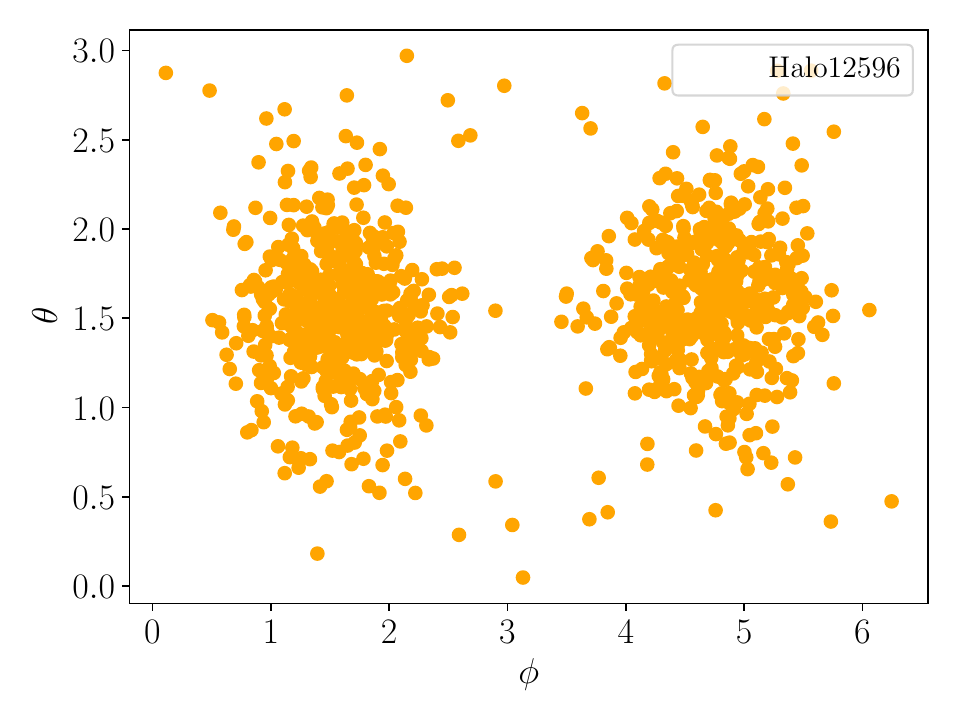}}
  \subfloat{
        \includegraphics[width=0.3\textwidth]{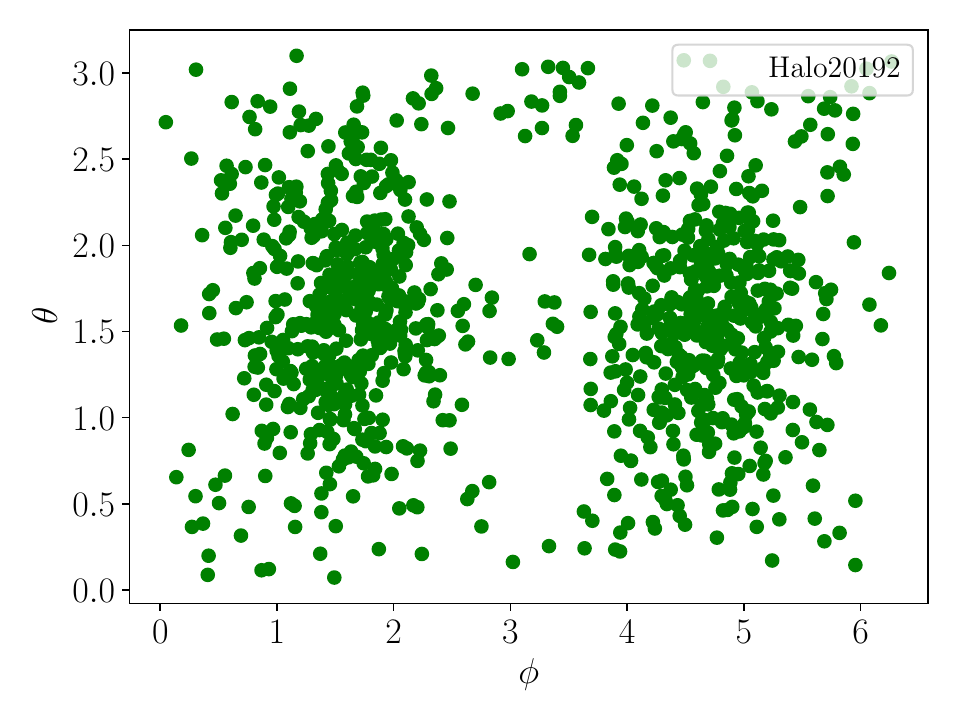}}
    ~ 
    \subfloat{
        \includegraphics[width=0.3\textwidth]{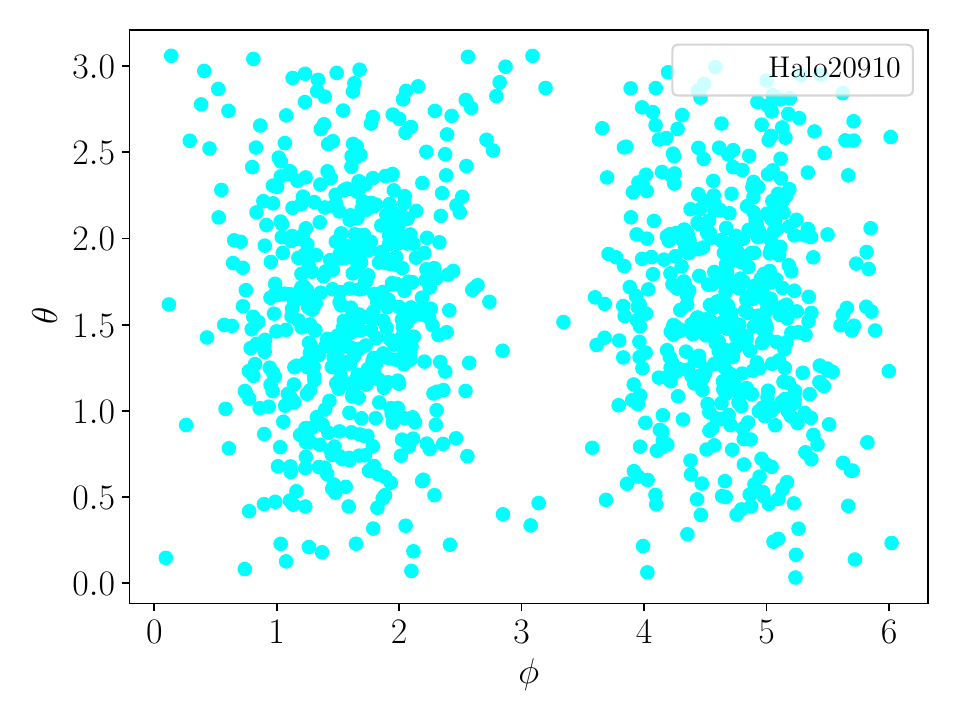}}\\
  \subfloat{
        \includegraphics[width=0.3\textwidth]{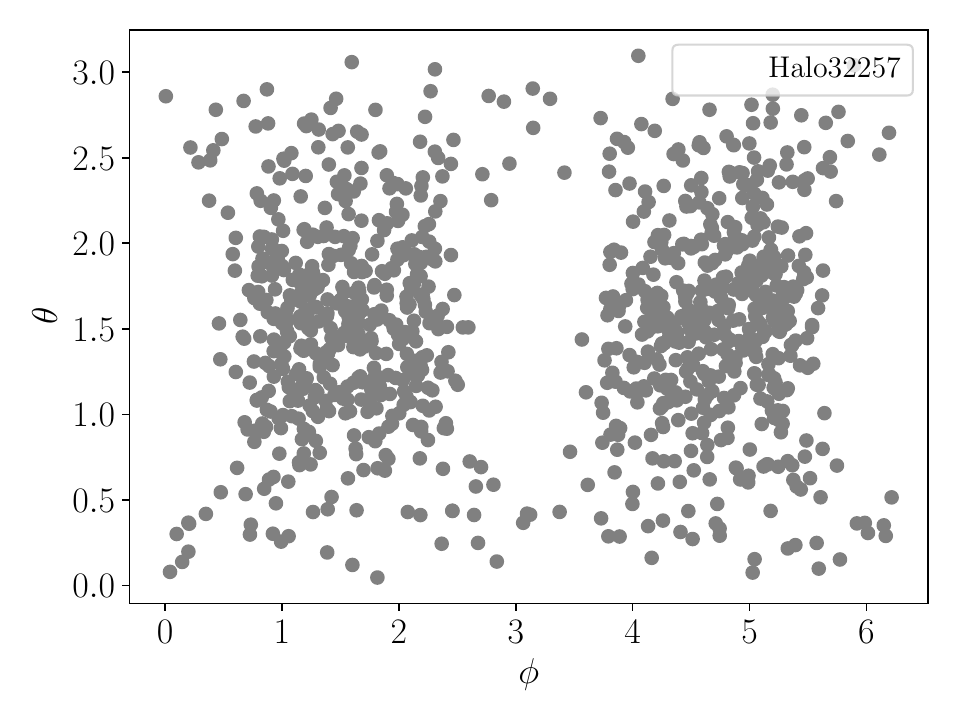}}
    ~ 
    \subfloat{
        \includegraphics[width=0.3\textwidth]{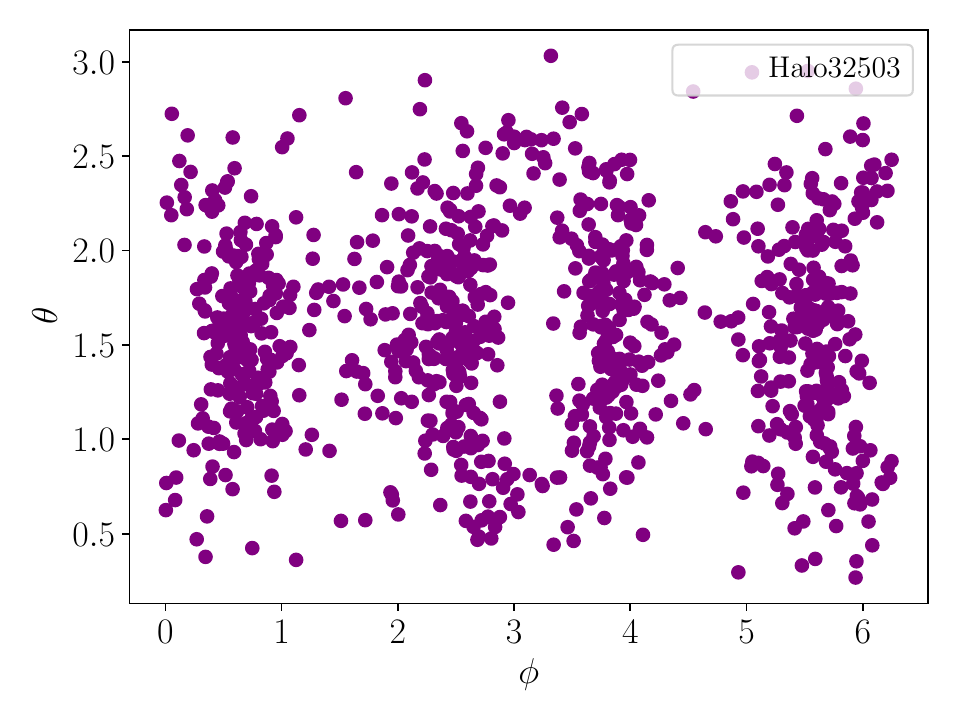}}
    \caption{Distribution of observation angles $(\phi, \theta)$ for each FIRE dwarf required to induce a correlation coefficient $r_{\epsilon \Sigma_*}$ and $r_{\epsilon \sigma_*}$ both $\geq 0.5$. As shown, the viewing angles of several galaxies must be highly tuned for this to be possible.}\label{fireangles}
\end{figure}

Relaxing further the assumption that all dwarf galaxies must draw from the same distribution of observing angles, we take random projections of all FIRE galaxies as before and record the projection angles of each if the resulting correlation coefficients  $r_{\epsilon \Sigma_*}$ and $r_{\epsilon \sigma_*}$ are both $\geq 0.5$. As shown in Fig~\ref{contour}, this is a rare event that occurs roughly once every $O(10^5)$ projections.  Fig.\ref{fireangles} shows the distribution of angles in $(\phi, \theta)$ space for each galaxy that allowed the collection to have a significant positive correlation. The more structure seen in each of these angle distributions, the more finely-tuned each galaxy's viewing angle must be to account for the correlation seen in the dim LG dsphs, and as shown for several galaxies the allowed region of projection angles is quite restrictive. Thus, as shown, it is unlikely that discrepancy between {dim LG and bright} FIRE galaxies can be explained by a non-uniformity of observing angles.

\section{Implications for Stellar and Dark Matter Morphology}\label{bary}

We have previously explained for measurements of many projections of a single galaxy, the measured (anti)-correlation between ellipticity and central surface brightness is determined by its morphology.  Here we show that the same is true for measurements of a single projection of a collection of distinct galaxies: we artificially introduce oblate morphologies by shifting the axis ratios of FIRE galaxies by a constant amount, and study the resulting shift in $r_{\epsilon \Sigma_*}$ expectation. While these do not represent realistic galaxy simulations, they provide one approximation to demonstrate that an oblate collection can produce the expected correlation shift, as simulations have not produced oblate halos to study. We show that these artificially oblate galaxies agree better with the dim LG dwarf population, and argue that these oblate stellar distributions may be sourced by oblate DM halos.

\begin{figure}[h!!]
    \centering
        \subfloat{\includegraphics[width=0.5\textwidth]{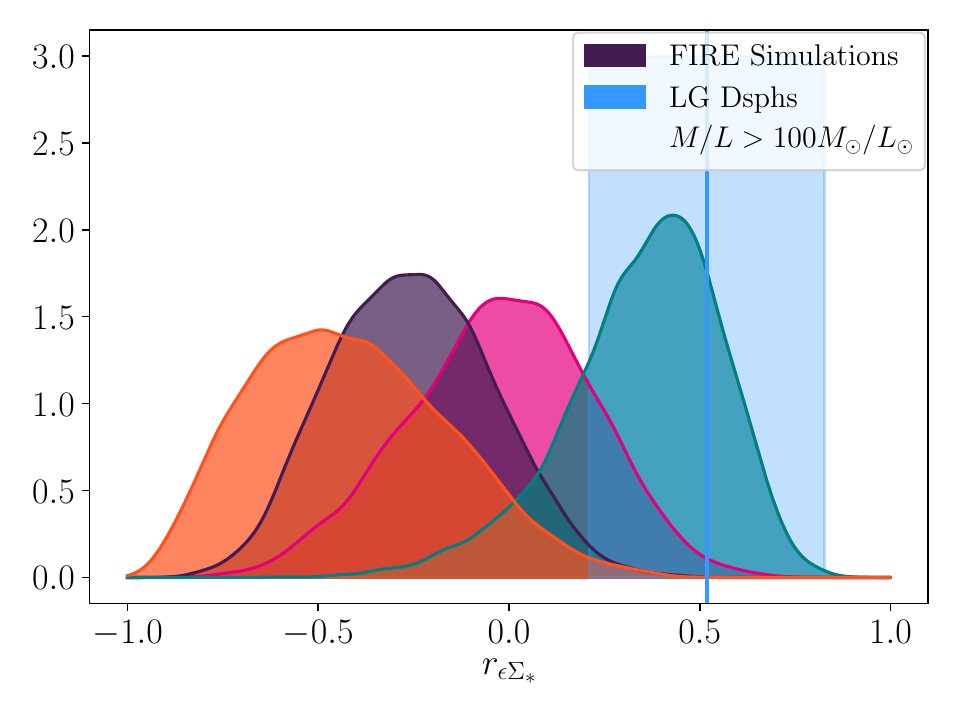}}
    ~ 
    \subfloat{\includegraphics[width=0.5\textwidth]{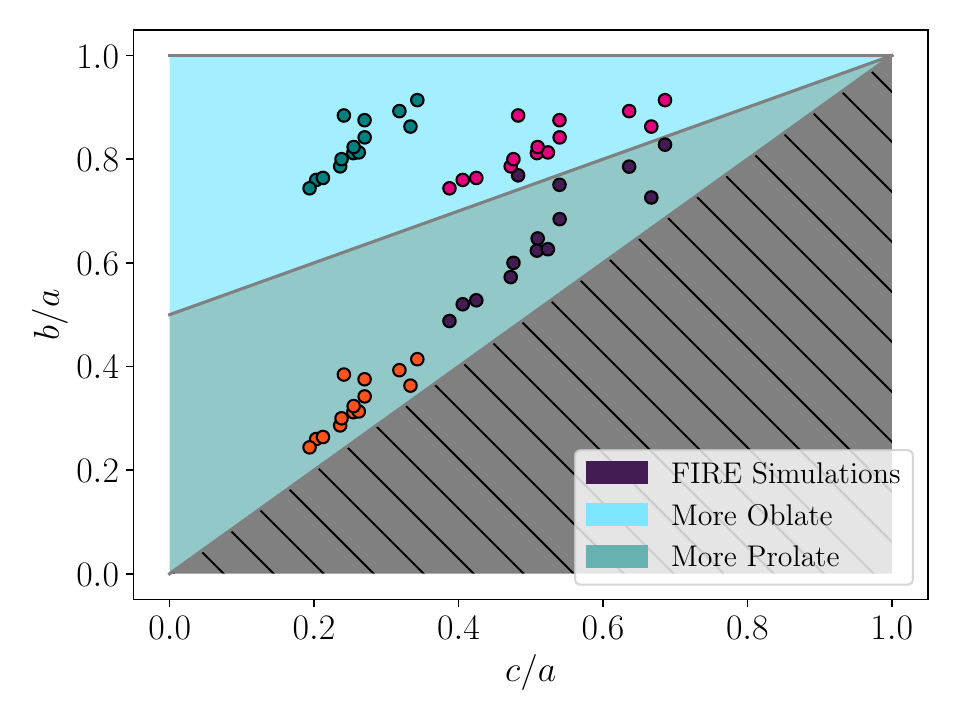}}
    \caption{Ellipticity-central surface brightness correlations for FIRE simulation galaxies with adjusted axis ratios -- the original morphologies (purple), $b/a$ shifted up (pink), $b/a$ shifted up and $c/a$ shifted down (blue), both $b/a$ and $c/a$ shifted down (orange)}-- and all other parameters fixed.  The modified axis ratios are shown on the right, and the induced correlation coefficient distribution via uniformly random projection is shown on the left with the same colors. The measured correlation for the dim dwarf sample is shown for reference. In both the original morphology is shown in dark purple. We observe that more oblate shapes do induce more positive correlation coefficients, even with an intrinsic dispersion in shapes and sizes.\label{axrat_tuning}
\end{figure}

Fig.~\ref{axrat_tuning} shows the resulting ellipticity-central surface brightness correlation distributions of $10^5$ random projections (left)  for shifted axis ratios (right), with all other parameters held fixed.  As shown,  a distribution of more oblate axis ratios (pink) will induce a more positive correlation between ellipticity and central surface brightness compared to the original morphologies (purple).  In addition to shifting into the oblate region (by increasing $b/a$), by shifting the axis ratios to be more elliptical (blue) -- i.e. with smaller $c/a$-- we obtain the strong positive correlation observed in the dim dwarf sample. In contrast, increasing the ellipticities while keeping the galaxies prolate (decreasing both $b/a$ and $c/a$, orange) will push the correlation-coefficient further into the negative regime.

As the population of dwarf spheroidals have been found to be generally isothermal and not significantly rotationally supported (\cite{ 2015ApJ...808..158B, annurev.astro.36.1.435}),  if the stellar population is in hydrostatic equilibrium then their morphology is set entirely by the gravitational potential.  In this case, the stellar distribution is set via the Jean's perscription 

 \begin{equation} \partial_i (\rho_{*} \sigma^2_{ij}) = - \rho_{*} \partial_{j} \Phi  \end{equation}

For the dim subpopulation where $\rho_{DM} \gg \rho_*$, it is valid to assume the potential is set entirely by the dark matter density by the Poisson equation  \begin{equation} \nabla^2 \Phi = 4\pi G \rho_{DM} \end{equation}

In fact, however, since these galaxies are the ones with the fewest baryons, it is likely that these objects have experienced tidal disruptions and it is not at all clear that they have reached a hydrostatic equilibrium.  Nevertheless, since these dim dwarf galaxies are some of the most dark matter dominated gravitational systems currently observable to us, it is interesting to consider what this ellipticity correlation discrepancy may imply for the shapes of these dark matter haloes. 

\begin{figure}[h!!]
\centering
\includegraphics[width=0.5\textwidth]{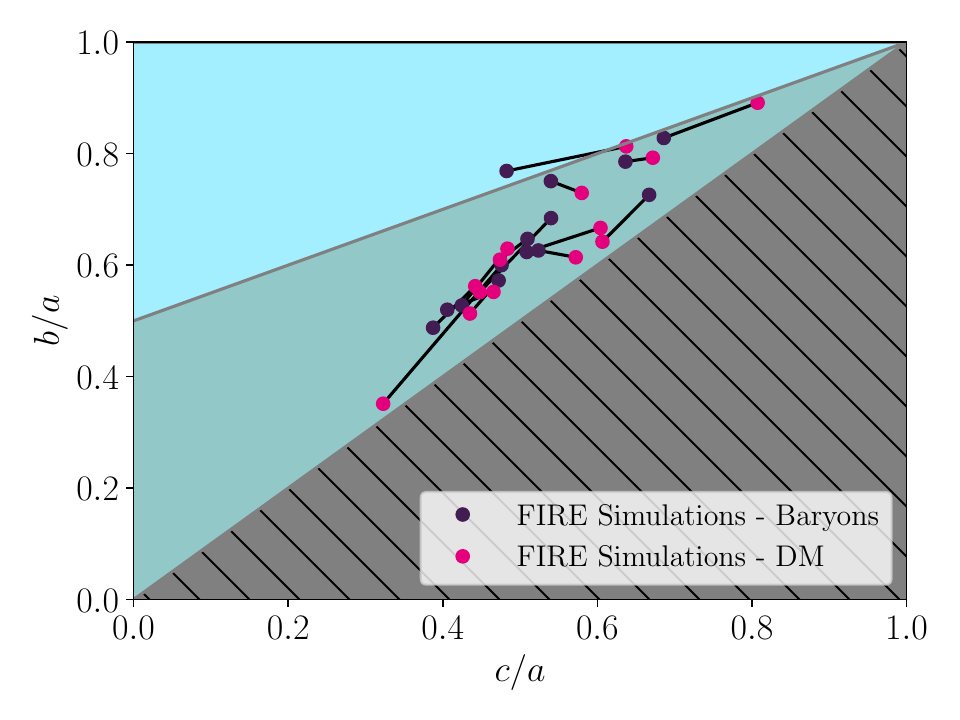}
\caption{Axis ratios of DM and Baryon distributions of the FIRE simulation galaxies. Corresponding DM haloes and stellar distributions are linked with a black line. While there doesn't seem to be a clean relationship between these two sets of axis ratios, both appear to be prolate and appear to lie within the same region of parameter space. A significantly more oblate population of baryon axis ratios would therefore necessitate a distinct population of DM halo morphologies.}\label{dmbary}
\end{figure}

Fig~\ref{dmbary} shows the axis ratios of the DM and stellar distributions of FIRE simulation dwarf galaxies. As shown, there does not appear to be a deterministic relationship between the dark matter and baryonic distributions, suggesting that many of these simulations may not have reached equilibrium, but nonetheless both are prolate with axis ratios lying in a similar region of parameter space. This indicates that the stellar population remains a reliable tracer for the underlying morphology of the dark matter halo, and simulations have yet to provide evidence for CDM halos that source oblate spheroidal galaxies.

In contrast, alternative models of dark matter may source oblate halos instead of prolate ones, and would likely predict oblate stellar distributions particularly in the low-stellar-mass regime. One such model is the Double-Disk Dark Matter (DDDM) model proposed in Ref. \cite{Fan:2013yva}. DDDM is a type of partially-interacting DM (PIDM) with a dominant cold component and a small strongly self-interacting component. The interacting component contains dark proton-like and electron-like particles which are allowed to exhibit dissipative dynamics and collapse to an angular-momentum supported thin dark disk, much like the baryonic disks of spiral galaxies.  Embedded in a CDM halo of small asphericity, the resulting dark matter halo will become oblate instead of prolate. 

Recent constraints using \textit{Gaia} data (\cite{Schutz:2017tfp}) show that a dark disk coplanar with the Milky Way disk can account for at most $\sim1\%$ of the MW halo mass. However, this result assumes an equilibrium distribution of stars, and it is not clear that this is true. In any case, this type of dark matter, if present, is expected to be found generically in all halos, and the presence of dark disks in dwarf galaxies is much less constrained. In these systems DDDM may induce an oblate DM halo, and the resulting oblate gravitational potential would cause larger ellipticity in the orbits of stars, ultimately sourcing an oblateness in the stellar component. In the case that these high mass-to-light dwarfs are indeed found to be oblate, this could provide motivation for a dark disk or similar DM models that generate oblateness in the stellar distribution without rotational support.



\section{Conclusions}\label{conc}

In this article we considered the correlations of observed ellipticity with stellar velocity dispersion as well as with central surface brightness of dwarf galaxies in and around the local group in order to infer their 3D morphologies and determine whether they are more likely to be oblate or prolate.  We find that dim LG dwarf galaxies (with $M/L > 100 M_\odot/L_\odot$) exhibit correlations of $r^{dim}_{\epsilon\Sigma_*} = 0.778 \pm 0.209 $ and $ r^{dim}_{\epsilon\sigma_*} =  0.832 \pm 0.323$, while bright LG dwarf galaxies (with $M/L < 100 M_\odot/L_\odot$) exhibit anti-correlation, with $r^{bright}_{\epsilon\Sigma_*} = -0.516 \pm 0.066$ and  $ r^{bright}_{\epsilon\sigma_*} = -0.153 \pm 0.117 $. This is consistent with oblate and prolate morphologies respectively. Assuming a uniform distribution of projection angles, we find that a set of 14 simulated galaxies from the FIRE project exhibit likewise an anti-correlation $r^{FIRE}_{\epsilon\Sigma_*} = -0.332\pm 0.213 $ and $ r^{FIRE}_{\epsilon\sigma_*} = -0.252\pm 0.214 $ consistent with generally prolate axis ratios. We find the discrepancy between simulated and dim LG dwarf galaxies (and between bright and dim LG dwarf galaxies themselves) cannot be explained by either subsampling the simulated or observed galaxies or relaxing the assumption of uniform projection angles, and the result holds for a bright-dim cut between $70-200 M_\odot$.  We propose that this discrepancy could  be explained by the dim LG dsphs exhibiting more oblate morphologies not currently captured by simulations, and a possible avenue of generating this oblateness is non-minimal interacting dark matter. As an example, we find that a $5\%$ fraction of radiative dark matter could induce a measureable change in ellipticity.  Our results do not preclude non-morphology-related avenues which would  result in a peak-like feature in both $\epsilon\Sigma_*$ and $\epsilon\sigma_*$ in LG dwarf galaxies, {or a CDM-compatible mechanism to generate pressure-supported oblateness in dim dwarf galaxies. A test of CDM consistency will lie in whether a corresponding feature can be found in a collection of simulated dim galaxies.}

Given the current estimators the LG dwarf galaxies with $M/L > 100 M_\odot/L_\odot$ appear to exhibit scaling expected for oblate bodies, and are morphologically distinct from both the brighter LG dwarf population and CDM hydrodynamic simulations which exhibit prolate-like behavior. Clearly, additional optical and spectral observations of particularly faint dwarf galaxies will improve the quality of dim LG dwarfs as an estimator for the underlying structure distribution. The next generation of sky surveys are poised to discover and characterize a collection of even dimmer dwarf galaxies, and it will be interesting see if the discrepancy we observe holds up against additional data.

With the data accessible to us, we have identified a discrepancy between current observations and the isolated dwarf galaxies from the FIRE simulation, but we are fundamentally interested in the correspondence between the Local Group dwarf galaxies and the $\Lambda$CDM-sourced expectation, and caveats remain on the comparability of the observed and simulated systems. In particular hydro-intensive simulations do not currently have the resolution to simulate host-satellite systems and the simulated galaxies we have used are isolated dwarfs, while observations are dominantly of satellites. Tidal disruptions, shocking, and ram pressure are expected to effect morphologies in ways that are not currently reflected in simulations. {In particular, it should be noted that this set of FIRE simulations do not provide a good representation of dim CDM galaxies, predominantly consisting of galaxies with $M/L < 100 M_\odot/L_\odot$.}  Further, these satellites are expected to be seeded by a range of halo masses (\cite{Brooks:2012vi}) while current FIRE simulations are all sourced by $\sim 10^{10} M_\odot$ halos. However, it is unclear that any of these tidal effects or an increased dark matter content will increase oblateness in galaxies, as larger CDM-only simulations predict prolate halos at all scales (\cite{2012AJ....144....4M}).   On the observation side, it is increasingly unclear that the model of a single isothermal stellar population is a valid description of dwarf galaxies (\cite{2012ApJ...756L...2A, 2018MNRAS.476.2168M, Battaglia:2006up}), and the galaxies observable to us are ultimately finite and preferentially luminous.    

It is useful to keep in mind that disagreements between these two samples currently could point to an inability of either the simulation or observation sample to represent its underlying distribution (due to insufficient sampling, imprecise measurements, etc.),  as well as indications of baryonic or dark matter physics that has yet to be modeled.  {The discrepancy we point out in this work could be due to any of these, and future studies on this will be informative as the quality of these datasets increase.} With consistent improvements in our ability to model hydrodynamics and to observe and identify ever fainter astrophysical objects, the systems that we simulate and observe are converging.

\section*{Acknowledgments}

We would like to thank Sasha Brownsberger, Andreas Burkert, Jakub Scholtz, and  Matt Walker for insightful discussions, as well as Mike Boylan-Kolchin and Alex Fitts for providing us with FIRE dwarf simulations. LR is supported by an NSF grant PHY-1620806, a Kavli Foundation grant ``Kavli Dream Team'', the Simons Foundation (511879),  Guggenheim and IHES CARMIN fellowships. LR also thanks Institute Henri Poincar\'e for their hospitality.

\bibliography{dark_matter_related} 
\bibliographystyle{aasjournal}

\end{document}